\documentclass[journal, 10pt, twocolumn, final]{IEEEtranTCOM}
\usepackage[T1]{fontenc}
\usepackage{cite}
\usepackage{nameref}
\usepackage{graphicx}
\usepackage{ifpdf,epsfig}
\usepackage{amssymb}
\usepackage[cmex10]{amsmath}
\usepackage{amsthm}
\usepackage{bm}
\usepackage{mdwmath}
\usepackage{mdwtab}
\usepackage[normalem]{ulem}
\usepackage{caption}
\usepackage{subcaption}
\usepackage[dvipsnames]{xcolor}
\usepackage{acro}
\DeclareAcronym{AWGN}{short = AWGN ,long = additive white Gaussian noise}
\DeclareAcronym{ACRDA}{short = ACRDA ,long = asynchronous contention resolution diversity ALOHA}
\DeclareAcronym{CDF}{short = CDF ,long = cumulative distribution function}
\DeclareAcronym{CRA-CC}{short = CRA-CC ,long = CRA-convolutional code}
\DeclareAcronym{CRA-SH}{short = CRA-SH ,long = CRA-shannon bound}
\DeclareAcronym{CRA}{short = CRA ,long = contention resolution ALOHA}
\DeclareAcronym{CRDSA}{short = CRDSA ,long = contention resolution diversity slotted ALOHA}
\DeclareAcronym{CRDSA++}{short = CRDSA++ ,long = contention resolution diversity slotted ALOHA++}
\DeclareAcronym{CRI}{short = CRI ,long = contention resolution interval}
\DeclareAcronym{CSA}{short = CSA ,long = coded slotted ALOHA}
\DeclareAcronym{CSI}{short = CSI ,long = channel state information}
\DeclareAcronym{DAMA}{short = DAMA ,long = demand assigned multiple access}
\DeclareAcronym{DSA}{short = DSA ,long = diversity slotted ALOHA}
\DeclareAcronym{DSSS}{short = DSSS ,long = direct sequence spread spectrum}
\DeclareAcronym{E_SSA}{short = E-SSA ,long = enhanced spread spectrum ALOHA}
\DeclareAcronym{ECRA}{short = ECRA ,long = enhanced contention resolution ALOHA}
\DeclareAcronym{ECRA-SC}{short = ECRA-SC ,long = ECRA selection combining}
\DeclareAcronym{ECRA-MRC}{short = ECRA-MRC ,long = ECRA maximal-ratio combining}
\DeclareAcronym{EGC}{short = EGC ,long = equal-gain combining}
\DeclareAcronym{FEC}{short = FEC ,long = forward error correction}
\DeclareAcronym{GEO}{short = GEO ,long = geostationary orbit}
\DeclareAcronym{IC}{short = IC ,long = interference cancellation}
\DeclareAcronym{IRCRA}{short = IRCRA ,long = irregular repetition contention resolution ALOHA}
\DeclareAcronym{IRSA}{short = IRSA ,long = irregular repetition slotted ALOHA}
\DeclareAcronym{LDPC}{short = LDPC ,long = low density parity check}
\DeclareAcronym{M2M}{short = M2M ,long = machine-to-machine}
\DeclareAcronym{MAC}{short = MAC ,long = medium access}
\DeclareAcronym{MF}{short = MF ,long = matched filter}
\DeclareAcronym{MF-TDMA}{short = MF-TDMA ,long = multi-frequency time division multiple access}
\DeclareAcronym{MRC}{short = MRC ,long = maximal-ratio combining}
\DeclareAcronym{MUD}{short = MUD ,long = multiuser detection}
\DeclareAcronym{PDF}{short = PDF ,long = probability density function}
\DeclareAcronym{PER}{short = PER ,long = packet error rate}
\DeclareAcronym{PLR}{short = PLR ,long = packet loss rate}
\DeclareAcronym{QPSK}{short = QPSK ,long = quadrature phase-shift keying}
\DeclareAcronym{RA}{short = RA ,long = random access}
\DeclareAcronym{RCB}{short = RCB ,long = random coding bound}
\DeclareAcronym{RTT}{short = RTT ,long = round trip time}
\DeclareAcronym{SA}{short = SA , long = slotted ALOHA}
\DeclareAcronym{SB}{short = SB ,long = Shannon bound}
\DeclareAcronym{SC}{short = SC ,long = selection combining}
\DeclareAcronym{SIC}{short = SIC ,long = successive interference cancellation}
\DeclareAcronym{SNIR}{short = SNIR ,long = signal-to-noise and interference ratio}
\DeclareAcronym{SINR}{short = SINR ,long = signal-to-interference and noise ratio}
\DeclareAcronym{SNR}{short = SNR ,long = signal-to-noise ratio}
\DeclareAcronym{TDMA}{short = TDMA ,long = time division multiple access}
\DeclareAcronym{UCP}{short = $\Code$-UCP ,long = $\Code$-unresolvable collision pattern}
\DeclareAcronym{VF}{short = VF ,long = virtual frame}
\begin{document}
\newtheorem{thm}{Theorem}
\newtheorem{lemma}{Lemma}
\newtheorem{coroll}{Corollary}
\newtheorem{prop}{Proposition}
\newtheorem{definition}{Definition}
\newtheorem{example}{Example}

\newcommand{\load}{\mathsf{G}}
\newcommand{\tp}{\mathsf{S}}
\newcommand{\se}{\xi}
\newcommand{\plr}{\mathsf{p}_l}
\newcommand{\rate}{\mathsf{R}}
\newcommand{\maxRate}{\mathsf{R}^*}
\newcommand{\mutInf}{\mathsf{I}}
\newcommand{\rvDec}{\mathcal{D}}

\newcommand{\capacity}{\mathsf{C}}
\newcommand{\capacitySC}{\capacity_\mathrm{S}}
\newcommand{\capacityMRC}{\capacity_\mathrm{M}}
\newcommand{\refCap}{\mathsf{C}_g}
\newcommand{\maxCap}{\mathsf{C}^*}
\newcommand{\maxSe}{\se^*}
\newcommand{\normCap}{\eta}

\newcommand{\dg}{\mathsf{d}}

\newcommand{\fraLen}{T_f}
\newcommand{\pkLen}{T_p}
\newcommand{\symLen}{T_s}
\newcommand{\fraTp}{n_p}
\newcommand{\VFStart}{T}
\newcommand{\RefTime}{t_0}
\newcommand{\Wind}{W}
\newcommand{\WindSh}{\Delta \Wind}

\newcommand{\numUs}{n_u}
\newcommand{\numSym}{n_s}
\newcommand{\numBit}{k}

\newcommand{\rx}{y}
\newcommand{\rxVec}{\bm{\rx}}
\newcommand{\tx}{x}
\newcommand{\txVec}{\bm{\tx}}
\newcommand{\noise}{n}
\newcommand{\noisePr}{\nu}
\newcommand{\noiseVec}{\bm{\noise}}
\newcommand{\noiseVar}{\sigma_{\noise}^2}
\newcommand{\intVal}{z}
\newcommand{\intVec}{\bm{\intVal}}
\newcommand{\intNum}{m}
\newcommand{\ch}{h}
\newcommand{\symVal}{a}
\newcommand{\pulse}{g}
\newcommand{\pulseRoot}{h}

\newcommand{\tm}{t}
\newcommand{\epoch}{\epsilon}
\newcommand{\freq}{f}
\newcommand{\phase}{\varphi}

\newcommand{\user}{u}
\newcommand{\replica}{r}
\newcommand{\sym}{s}

\newcommand{\usPw}{\mathsf{P}}
\newcommand{\PwAg}{\usPw_g}
\newcommand{\usPwTx}{\usPw_t}
\newcommand{\noisePw}{\mathsf{N}}
\newcommand{\noiseSD}{N_0}
\newcommand{\intPw}{\mathsf{Z}}
\newcommand{\sinr}{\gamma}
\newcommand{\sinrVec}{\Gamma}

\newcommand{\UCP}{\mathcal{L}}
\newcommand{\UCPcons}{\mathcal{L}^*}
\newcommand{\AllUCP}{\mathcal{L}_\mathrm{S}}
\newcommand{\Code}{\mathcal{C}}
\newcommand{\CollClus}{\mathcal{S}}

\newcommand{\RStart}{\tau}
\newcommand{\VL}{\RStart_l^*}
\newcommand{\VR}{\RStart_r^*}
\newcommand{\Vg}{\RStart^*}
\newcommand{\Vpd}{T_v}
\newcommand{\nVp}{n_v}

\newcommand{\mU}{\alpha_u}
\newcommand{\sL}{\beta_d}
\newcommand{\UL}{\beta_{u-d}}

\newcommand{\intFreeAsy}{\varphi_a}
\newcommand{\intFreeMRC}{\varphi_m}
\newcommand{\intOne}{\mu}
\newcommand{\rateFree}{\rate_f}
\newcommand{\rateInt}{\rate_i}
\newcommand{\rateIntO}{\rate_{i1}}
\newcommand{\rateIntT}{\rate_{i2}}

%%%%% Color for Editorial Changes
\definecolor{gl}{rgb}{0.0,0.5,0.8}
\definecolor{fc}{rgb}{0.8,0.5,0}
\definecolor{al}{rgb}{1,0.3,0.3}
\newcommand{\giangio}{\textcolor{gl}}
\newcommand{\fede}{\textcolor{fc}}
\newcommand{\alert}{\textcolor{al}}

% \input{./sections/displaynotation}
\title{Enhancing Contention Resolution ALOHA\\ using Combining Techniques}

%\author{
%\IEEEauthorblockN{Federico Clazzer\IEEEauthorrefmark{1}, Christian Kissling\IEEEauthorrefmark{2}, Mario Marchese\IEEEauthorrefmark{3}}\\
%\IEEEauthorblockA{\IEEEauthorrefmark{1}German Aerospace Center (DLR), Institute of Communications and Navigation\\ 82234 Oberpfaffenhofen, Germany. E-mail: federico.clazzer@dlr.de}\\
%\IEEEauthorblockA{\IEEEauthorrefmark{2}Munich University of Applied Sciences, Department for Electrical Engineering and Information Technology, 80335 Munich, Germany.\\E-mail: christian.kissling@hm.edu}\\
%\IEEEauthorblockA{\IEEEauthorrefmark{3}University of Genoa, Department of Electrical, Electronic, and Telecommunications Engineering, and Naval Architecture (DITEN), 16145 Genova, Italy. E-mail: mario.marchese@unige.it}
%}

\author{Federico~Clazzer,~\IEEEmembership{Student Member, IEEE,}
        Christian~Kissling,
        Mario~Marchese,~\IEEEmembership{Senior Member, IEEE}%
\thanks{%
The material in this paper was presented in part at the 2013 International ITG Conference on Systems, Communications and Coding (SCC).}%
\thanks{
Federico Clazzer is with the Institute of Communications and Navigation of the German Aerospace Center (DLR), Muenchner Strasse 20, 82234 Wessling, Germany. Email: \texttt{Federico.Clazzer@dlr.de.}}%
\thanks{
Christian Kissling is with the Department for Electrical Engineering and Information Technology of the Munich University of Applied Sciences, {Lothstr. 64}, 80335 Munich, Germany. E-mail: \texttt{Christian.Kissling@hm.edu}.}%
\thanks{
Mario Marchese is with the DITEN, University of Genoa, {Via Opera Pia 13}, 16145 Genova, Italy. E-mail: \texttt{Mario.Marchese@unige.it.}
\newline
\newline
\textcopyright 2017 IEEE. Personal use of this material is permitted. Permission from IEEE must be obtained for all other uses, in any current or future media, including reprinting/republishing this material for advertising or promotional purposes, creating new collective works, for resale or redistribution to servers or lists, or reuse of any copyrighted component of this work in other works.\newline Digital Object Identifier 10.1109/TCOMM.2017.2759264.}
%\thanks{Corresponding Address: Federico Clazzer, KN-SAN, DLR, Muenchner Strasse 20, 82234 Wessling, Germany. Tel: +49-8153 28-1120, Fax: +49-8153 28-2844, E-mail: \texttt{federico.clazzer@dlr.de}.}
}

%\markboth{submitted to IEEE Transactions on Wireless Communications}{F. Clazzer et al.: Enhancing Contention Resolution ALOHA using Combining Techniques}%


\maketitle

\begin{abstract}
%\boldmath
Recently, random access protocols have acquired a new wave of interest, not only from the satellite communication community, but also from researchers active in fields like Internet of Things and machine-to-machine. Asynchronous (slot- and frame-wise) ALOHA-like random access protocols, are very attractive for such applications, enabling low complexity transmitters and avoiding time synchronization requirements. Evolutions of ALOHA employ time diversity through proactive replication of packets, but the time diversity is not fully exploited at the receiver. Combining techniques, as selection combining and maximal-ratio combining, are beneficial and are adopted in the \ac{ECRA} scheme, presented here. A tight approximation of the packet loss rate for asynchronous random access, including \ac{ECRA}, well suited for the low channel load region is derived. Finally, \ac{ECRA} is evaluated in terms of spectral efficiency, throughput and packet loss rate in comparison with recent protocols, showing that it is able to largely outperform both slotted synchronous and asynchronous schemes.
\end{abstract}


% \begin{IEEEkeywords}
% Random Access, Successive Interference Cancellation, Combining Techniques, Collision Resolution, Asynchronous Random Access, Medium Access.
% \end{IEEEkeywords}

% \fi
\IEEEpeerreviewmaketitle

% Different paper sections

\section{Introduction}
\acresetall

\IEEEPARstart{F}{ixed} allocation, for multiple-access wireless communication systems that are dynamic in terms of resource requests from the transmitting nodes, is normally inefficient. On demand resource allocation, e.g. \ac{DAMA}, can be beneficial \cite{Abramson1994}. However, there are a number of scenarios where \ac{DAMA} is unable to counteract efficiently the dynamics of the resource requests. For example, when the channel traffic shows a bursty and unpredictable nature, when there is a very high number of transmitters and coordination is hard to achieve, or when delay-critical applications are considered. The last case is of particular importance in \ac{GEO} satellite communication systems. Before the resource request of a given transmitter can be satisfied, an additional delay of one round trip time of ca. $500$ ms may elapse. Such additional delay is critical for several applications \cite{Kissling2011a}.

% state-of-the-art
\begin{sloppypar}
\Ac{RA} protocols have evolved significantly from the original idea of ALOHA proposed by Abramson in 1970 \cite{Abramson1970}, and its time slotted evolution \cite{Roberts1975, Abramson1977}. Recently, driven by a number of applications like underwater networks \cite{Pompili2009}, RFID communication systems \cite{He2013}, vehicular ad hoc networks \cite{Menouar2006}, \ac{M2M} communication systems \cite{Stefanovic2013TCOM} and satellite networks \cite{Peyravi1999}, numerous new \ac{RA} schemes have been proposed. Among them, worth to be mentioned is \ac{CRDSA} \cite{Casini2007}. The scheme is an evolution of \ac{DSA} \cite{Choudhury1983}. With respect to \ac{SA}, in \ac{DSA}, lower delays and higher throughput are provided under very moderate channel load conditions, by transmitting in a proactive way two or more times the same packet. The packet instances (replicas in the following) are sent separately with a random delay. Nonetheless, the increase of channel occupation, due to the presence of replicas, is counterproductive for medium channel load conditions, and worsen both throughput and \ac{PLR} performance w.r.t. \ac{SA}. In \ac{DSA} arises the tradeoff between time diversity introduced by the presence of replicas, and the channel load increase. When the channel load is limited, replicating packets is beneficial because the collision probability remains relatively low while the probability that at least one replica is received collision-free is higher. Instead, at moderate channel load, the \ac{PLR} is driven by collisions, and sending replicas harms the performance.
\end{sloppypar}

A larger gain is achieved when both time diversity and \ac{SIC} are exploited \cite{Casini2007}. The \ac{CRDSA} scheme \cite{Casini2007} follows the idea of \ac{DSA} to send more than one replica per user, and additionally introduces \ac{SIC} at the receiver. Transmissions are organized into frames, where users are allowed to transmit only once. The users replicate their packets two (or more) times, and place the replicas in slots selected uniformly at random, providing in all replicas the information on the selected slots. At the receiver, \ac{SIC} exploits the presence of multiple replicas per user for clearing up collisions. Every time a packet is decoded, \ac{SIC} reconstructs the waveform and subtracts it from all the slot locations selected for transmission by the corresponding user, possibly removing the interference contribution with respect to other packets. The performance evaluations in \cite{Casini2007} have shown that the maximum throughput of \ac{CRDSA} can be impressively extended from $\tp\cong 0.36$ (the peak throughput of \ac{SA}, measured in average number of successful transmissions per transmission period \cite{Kleinrock1976_book} or packets per slot\footnote{Following the definition of \cite{Kleinrock1976_book}, we assume that a transmission period is equal to $\pkLen$ seconds, which coincides with the physical layer packet duration and also coincides with the slot duration. Therefore, for slotted protocols, the throughput can be measured also in packets/slot.}), up to $\tp\cong 0.55$. For a target \ac{PLR} of $10^{-3}$, \ac{CRDSA} can support up to $1$ $\mathrm{[bits/symbol]}$, considering \ac{QPSK} modulation and the 3GPP Turbo Code of rate $1/3$ \cite{Herrero2014}. Further performance improvements can be achieved when, 1) more than two replicas per user and per frame are sent, 2) difference in received power, due to induced power unbalance or fading, and capture effect are considered \cite{Herrero2014}. The stability of \ac{CRDSA} has been investigated in \cite{Kissling2011}, while more recently an analytical framework for slotted \ac{RA} protocols embracing \ac{SA}, \ac{DSA} and \ac{CRDSA} has been presented in \cite{Herrero2014}. \Ac{IRSA} \cite{Liva2011} is an extension of \ac{CRDSA}, where the number of replicas sent by users is drawn from a probability mass function optimized for maximizing the throughput. The bipartite-graph representation is introduced, and exploited for characterizing the \ac{IC} process, helping the optimization procedure, and building a bridge towards tools typically deployed in coding theory. An extension of \ac{IRSA}, named \ac{CSA} has been presented and analysed in \cite{Paolini2014}, where the replicas are not simply repetitions of the original packet as in \ac{IRSA} and \ac{CRDSA}, but instead, they are coded versions of them. \ac{IRSA} approaches a theoretical throughput of $\tp=0.97$ with a distribution containing a maximum of 16 replicas per user, obtained via differential evolution \cite{Liva2011}. Both \ac{IRSA} and \ac{CSA} are able to achieve a throughput arbitrarily close to 1 packet per slot \cite{Narayanan2012}, under the collision channel model \cite{Tong_2004}, letting both the number of slots in the frame, and the maximum number of replicas sent by each user, grow very large. %\ROne{For a frame composed by $200$ slots and a maximum of $8$ replicas, \ac{IRSA} can support channel load up to $0.75$ not exceeding the \ac{PLR} of $10^{-2}$, \cite{Liva2011}.}
The authors in \cite{Bui2015}, bridge soft combining with \ac{SIC} for the \ac{CRDSA} scheme, showing remarkable gains. In \cite{Gamb_Schlegel2013} it has been shown that joint decoding of the collided packets can be attempted, resorting to \ac{MUD} techniques. The authors of \cite{Stefanovic2012,Stefanovic2013TCOM} elaborate the concept of \emph{frameless} slotted scheme, i.e. the duration of a frame is not a-priori fixed, but the contention ends when the throughput is maximized. Further evolutions of \ac{RA} include the extension to multiple receiver scenarios \cite{Munari2013,Jakovetic2015} and to all-to-all broadcast transmission \cite{Ivanonv2015}.

%Relevant for the present work is the idea proposed by the authors in \cite{ZigZag} to adopt an iterative decoding procedure in order to counteract the hidden terminal problem in the IEEE $802.11$ \ac{MAC} protocol. The authors observe that packets colliding once are very likely to collide again in retransmissions. However, the jitter may differ in the two collisions and can be exploited for triggering a decoding procedure that identifies portions of the packets free from interference, and proceeds eliminating the demodulated portion into the other collision. This can free from interference a section of the second packet collided, that can be now decoded and removed from the first collision. Iterating the procedure can possibly lead to decode both packets. Extension to this approach are proposed in \cite{SigSag} where instead of the hard-decoded symbols, soft-information is propagated.

% knowledge gaps
Similarly, \ac{SIC} can also be adopted when considering asynchronous spread spectrum random access, as in \cite{delRioHerrero_2012}. The \ac{E_SSA} uses direct sequence spread spectrum at the transmitter without replicas, i.e. each terminal sends only one packet per transmission. At the receiver side, \ac{SIC} is employed for removing interference, once packets are correctly decoded. The excellent performance of \ac{E_SSA} are shown in \cite{Gallinaro_2014}.

For asynchronous \ac{RA} without spreading, \ac{CRA} \cite{Kissling2011a} has been the first attempt to mimic the improvements given by \ac{CRDSA}. Time slots are removed, but frames are kept, and users are allowed to transmit their replicas within the frame, without any constraint excepts avoiding self-interference. At the receiver, \ac{SIC} is employed to improve the performance, similarly to the slotted counterpart \ac{CRDSA}. Recently \ac{ACRDA} \cite{DeGaudenzi2014} has removed also the frame structure still present in \ac{CRA}, reducing once more the transmitter complexity. However, both \ac{CRA} and \ac{ACRDA} do not exploit the inherent time diversity of the interference among replicas, which naturally arises due to the asynchronous nature of the protocol, i.e. different portion of replicas of a given user might be interfered.

% objectives
Driven by this observation, the present paper introduces the \ac{ECRA} slot- and frame-wise asynchronous \ac{RA} scheme. It employs combining techniques in order to resolve collision patterns where \ac{SIC} alone is unable to succeed. The main contributions of the present work can be summarized as:
 \begin{itemize}
 \item{\emph{Extension of asynchronous \ac{RA} protocols towards combining techniques} such as \ac{SC}, \ac{EGC} and \ac{MRC} \cite{Brennan1959,Jakes1974}. The novel \ac{ECRA} exploits time diversity of the interference pattern suffered by the replicas, for creating a combined observation at the receiver, on which decoding is attempted.
 \item{\emph{Development of an analytical approximation of the \ac{PLR} performance, for asynchronous \ac{RA} schemes}, particularly tight for low channel load. The approximation focuses on a subset of collision patterns, unresolvable with \ac{SIC}. The \ac{PLR} analytical approximation for \ac{ECRA} with \ac{MRC} focuses on the case with two replicas only.}
 \item{\emph{Comparison of \ac{ECRA} with asynchronous and slot synchronous protocols under several metrics} as throughput, spectral efficiency and normalized capacity.}}
 \end{itemize}

The reminder of the paper is organized as follows. In Section \ref{sec:sys_ov}, the system model is presented, including the detailed description of the \ac{ECRA} decoding algorithm. In Section \ref{sec:PLR}, the \ac{PLR} approximation is derived for asynchronous schemes, including \ac{ECRA}. In Section \ref{sec:simulations_all}, the performance metrics are defined, and numerical results for \ac{ECRA} as well as comparison with recent slot synchronous and asynchronous schemes are shown. Finally, in Section \ref{sec:conclusions}, concluding remarks close the paper.

%The rest of the paper is organized as follows. Next subsection will make a general literature review. In section II, the system considered is detailed, while in section III the loop definition is made and the interference characteristic of an Aloha-like protocol discussed. Furthermore, a lower bound on the number of loops per \ac{MAC} frame is derived in the same section. In section IV the decoding procedure of \ac{ECRA} in both its variants is presented. In section V the performance metric definition is carried out, while in section VI numerical results for \ac{ECRA} protocol compared with the state-of-the-art protocol are presented. Finally in section VII the conclusions are drawn.

\section{System Model}
\label{sec:sys_ov}
We assume an infinite user population generating traffic, following a Poisson process of intensity $\load$. The channel load\footnote{The channel load corresponds to the \emph{logical load} $\load$, since it takes into consideration the net information transmitted, depurated from the number of replicas per user $\dg$.} $\load$ is measured in packet arrivals per packet duration $\pkLen$. Upon arrival, each user replicates its packet $\dg$ times, with $\dg$ the repetition degree of the system. The first replica is transmitted immediately, while the remaining $\dg-1$ are sent within a \ac{VF} of duration $\fraLen$, starting at the beginning of the first replica.\footnote{It is important to underline that, the concept of \ac{VF} has been firstly introduced in \ac{ACRDA} \cite{DeGaudenzi2014}, and was not present neither in \ac{CRA} nor in the first statement of \ac{ECRA} \cite{Clazzer2012}.} As a consequence, virtual frames are asynchronous among users. Replicas are sent such that self-interference is avoided. The time location within the \ac{VF} of each replica is stored in a dedicated portion of the packet header. Each replica is composed by $\numBit$ information bits. In order to protect the packets against channel impairments and interference, a channel code $\Code$ with Gaussian codebook is adopted. We define the coding rate $\rate=\numBit/\numSym$, where $\numSym$ is the number of symbols within each packet after channel encoding and modulation. We denote with $\symLen$ the duration of a symbol so that $\pkLen=\symLen \numSym$. Replicas are then transmitted through an \ac{AWGN} channel.

Let us consider the transmitted signal $\tx^{(\user)}$ of the $\user$-th user,
\begin{equation}
\label{eq:tx_sig}
\tx^{(\user)}(\tm)= \sum_{i=0}^{\numSym-1} \symVal_i^{(\user)} \pulse(\tm-i \symLen).
\end{equation}
Where $\bm{\symVal}^{(\user)}=\left( \symVal_0^{(\user)}, \symVal_1^{(\user)}, \dots, \symVal_{\numSym-1}^{(\user)} \right)$ is the codeword of user $\user$ and $\pulse(\tm)=\mathcal{F}^{-1}\left\{\sqrt{\mathtt{CR}(f)}\right\}$ is the pulse shape, being $\mathtt{CR}(f)$ the frequency response of the raised cosine filter. The generic user $\user$ signal is affected by a frequency offset, modeled as an uniformly distributed random variable $\freq^{(\user)} \sim \mathcal{U}\left[-f_{\mathrm{max}};f_{\mathrm{max}}\right]$, and a sampling epoch (cfn. \cite{Mengali_Dandrea} Chapter $2$), also modeled as an uniformly distributed random variable $\epoch^{(u)} \sim \mathcal{U}\left[0;T_s\right)$. Both frequency offset and sampling epoch are common to each replica of the same user, but independent user by user. The phase offset is modeled as a random variable uniformly distributed between $0$ and $2\pi$, i.e. $\phase^{(\user,\replica)} \sim \mathcal{U}\left[0;2\pi\right)$, and it is assumed to be independent replica by replica. Assuming that $f_{\mathrm{max}}T_s\ll 1$, the received signal $\rx(\tm)$, after matched filtering, can be approximated as
\ifCLASSOPTIONtwocolumn
    \begin{equation}
    \begin{aligned}
    \label{eq:rx_sig}
    \rx(\tm) &\cong \sum_{\user} \sum_{\replica=0}^{\dg-1} \tilde{\tx}^{(\user)}(\tm - \epoch^{(u)} - \VFStart^{(\user,\replica)} - \RefTime^{(\user)}) e^{j\left(2\pi \freq^{(\user)} + \phase^{(\user,\replica)}\right)}\\ &+ \noise(\tm)
    \end{aligned}
    \end{equation}
\else%
    \begin{equation}
    \label{eq:rx_sig}
    \rx(\tm) \cong \sum_{\user} \sum_{\replica=0}^{\dg-1} \tilde{\tx}^{(\user)}(\tm - \epoch^{(u)} - \VFStart^{(\user,\replica)} - \RefTime^{(\user)}) e^{j\left(2\pi \freq^{(\user)} + \phase^{(\user,\replica)}\right)} + \noise(\tm)
    \end{equation}
\fi
\begin{sloppypar}
with $\tilde{\tx}^{(\user)}=\sum_{i=0}^{\numSym-1} \symVal_i^{(\user)} \tilde{\pulse}(\tm-i \symLen)$, where ${\tilde{\pulse}(\tm) = \mathcal{F}^{-1}\left\{\mathtt{CR}(f)\right\}}$. In equation \eqref{eq:rx_sig}, $\VFStart^{(\user,\replica)}$ is the delay w.r.t. the \ac{VF} frame start for user $\user$ and replica $\replica$, while $\RefTime^{(\user)}$ is the $\user$-th user delay w.r.t. the common reference time. The noise term $\noise(\tm)$ is given by $\noise(\tm) \triangleq \noisePr(\tm) \ast \pulseRoot(\tm)$, where $\noisePr(\tm)$ is a white Gaussian process with single-sided power spectral density $\noiseSD$ and $\pulseRoot(\tm)$ is the \ac{MF} impulse response of the root raised cosine filter, i.e. $\pulseRoot(\tm) = \mathcal{F}^{-1} \left\{ \sqrt{\mathtt{CR}(f)} \right\}$.
\end{sloppypar}

\begin{sloppypar}
For the $\user$-th user, $\replica$-th replica, assuming an ideal estimate of the sampling epoch $\epoch^{(u)}$, the frequency offset $\freq^{(\user)}$ and the phase offset $\phase^{(\user,\replica)}$, the discrete-time version of the received signal ${\rxVec^{(\user,\replica)} = (\rx_0^{(\user,\replica)}, ..., \rx_{\numSym-1}^{(\user,\replica)})}$ is given by
\begin{equation}
\rxVec^{(\user,\replica)} = \bm{\symVal}^{(\user)} + \intVec^{(\user,\replica)} + \noiseVec.
\end{equation}
Here %$\txVec^{(\user)} = \bm{\symVal}^{(\user)}$,
$\intVec^{(\user,\replica)}$ is the interference contribution over the user-$\user$ replica-$\replica$ signal and $\noiseVec=(\noise_0,...,\noise_{\numSym-1})$ are the samples of a complex discrete white Gaussian process with ${\noise_i \sim \mathcal{CN}(0,2\noiseVar)}$.
\end{sloppypar}

The instantaneous \ac{SINR} $\sinr$ for the $i$-th sample of the $\user$-th user $\replica$-th replica is
\begin{equation}
\sinr_{i}^{(\user, \replica)} = \frac{\usPw_{i}^{(\user)}}{\noisePw + \intPw_{i}^{(\user, \replica)}}
\end{equation}
with $\usPw_{i}^{(\user)} \triangleq \mathbb{E}\left[|\symVal_{i}^{(\user)}|^2 \right]$, $\noisePw=2\noiseVar$ and $\intPw_{i}^{(\user, \replica)} \triangleq \mathbb{E}\left[ |\intVal_{i}^{(\user,\replica)}|^2 \right]$, which is the aggregate interference power contribution on the $i$-th sample of the considered replica. Throughout the paper, we assume that all users are received with the same power, i.e. perfect power control is adopted. Hence, $\usPw_{i}^{(\user)} = \usPw$ and $\intPw_{i}^{(\user, \replica)} = \intNum_i^{(\user, \replica)} \usPw$, where $\intNum_i^{(\user, \replica)}$ denotes the number of active interferers over the $i$-th symbol of the $\user$-th user $\replica$-th replica. The aggregate interference is a discrete Gaussian process, with $\intVal_{i} \sim \mathcal{CN}\left(0,\intNum_i^{(\user, \replica)} \usPw\right)$, and the \ac{SINR} thus becomes
\begin{equation}
\sinr_{i}^{(\user, \replica)} = \frac{\usPw}{\noisePw + \intNum_i^{(\user, \replica)} \usPw}.
\end{equation}

\begin{sloppypar}
The \ac{SINR} vector over the $\numSym$ symbols of the considered replica is denoted with ${\sinrVec^{(\user, \replica)} = (\sinr_{0}^{(\user, \replica)}, \sinr_{1}^{(\user, \replica)}, ..., \sinr_{\numSym-1}^{(\user, \replica)})}$.
\end{sloppypar}

\subsection{Modeling of the Decoding Process}
\label{sec:int_model}

Typically, the destructive collision channel model is adopted \cite{Tong_2004} in the analysis of the \ac{MAC} layer of \ac{RA} protocols. This physical layer abstraction assumes that, only packets received collision-free can be correctly decoded, while all packets involved in collisions are lost. This assumption is, in general, inaccurate when packets are protected with an error correcting code, and for asynchronous schemes specifically, is particularly pessimistic.\footnote{Also for slot synchronous RA with powerful error correcting codes, decoding of packets may be possible even in presence of interference.} In fact, interference can be counteracted by the error correction code, and some collisions can be resolved.

Motivated by this, we resort to a \emph{block interference model} \cite{McEliece1984} given by $\numSym$ subsequent Gaussian channels \cite{cover2006} (one for each replica symbol), where the $i$-th channel is characterized by a \ac{SNR} $\sinr_{i}$.\footnote{We are omitting here the superscript $(\user, \replica)$ for ease of notation.} Similarly to \cite{Thomas_2000}, the idea is to take into account the mutual information carried by each replica symbol, and then compute the average over the entire replica. Leveraging on the Gaussian assumption of both the signals and noise,\footnote{We shall point out that, under some specific conditions, the Gaussian assumption for the interference can also be a good approximation for linear modulated and Turbo encoded signals. See e.g. \cite{Herrero2014}.} the instantaneous mutual information over the $i$-th channel $\mutInf(\sinr_{i})$ is
\begin{equation}
\mutInf(\sinr_{i}) = \log_2( 1 + \sinr_{i} ).
\end{equation}
%\ROne{As shown in \cite{Herrero2014}, even in the worst case scenario of slot- and symbol-synchronous colliding packets with equal power, the resulting interference can be tightly approximated with a Gaussian process. This is true also for collision between two packets only.}
Differently from the classical parallel Gaussian channel problem of finding the best power allocation per channel, in order to maximize capacity (cf. Chapter 10.4 of \cite{cover2006}), here \ac{CSI} is not present at the transmitter, since the interference contribution cannot be predicted due to the uncoordinated user transmissions. Therefore, the power allocation over the channels, i.e. symbols of the replica, is kept constant and is not subject to optimization. The instantaneous mutual information, averaged over the $\numSym$ channels, is
\begin{equation}
\label{eq:mi_dec}
\mutInf(\sinrVec) = \frac{1}{\numSym} \sum_{i=0}^{\numSym-1} \mutInf(\sinr_{i}) = \frac{1}{\numSym} \sum_{i=0}^{\numSym-1} \log_2( 1 + \sinr_{i} ).
\end{equation}
The interference has been modeled similarly in \cite{Thomas_2000}. We introduce a binary variable $\rvDec$, modelling the decoding process, such that
\begin{equation}
\begin{split}
\rvDec &= 1 \qquad \text{if decoding succeeds}\\
\rvDec &= 0 \qquad \text{otherwise}.
\end{split}
\end{equation}
We have
\begin{equation}
\label{eq:dec_res}
\rvDec = \mathbb{I}\left\{\rate \leq \mutInf(\sinrVec)\right\}
%\rvDec = 1\left\{\rate \leq \mutInf(\sinrVec)\right\}
\end{equation}
where $\mathbb{I}\{X\}$ denotes the indicator (Inverson) function.\footnote{This model allows to take into account features like channel coding, multi-packet reception and capture effect \cite{Ghez88:MPR, Zorzi94:CaptureAloha}.} Observe that, the destructive collision model is a special case, where the rate $\rate$ is chosen such that only packets collision-free can be succesfully decoded, i.e. $\rate = \log_2\left(1+\frac{\usPw}{\noisePw}\right)$. The decoding process model based on the threshold induced by the selected rate, has some non-negligible effect on the performance with respect to more accurate models, that take into account the specific channel code and block length. Nevertheless, it is a good first approximation for highlighting the improvements given by the proposed scheme.

\subsection{Enhanced Contention Resolution ALOHA Decoding Algorithm}
\begin{figure*}
\centering
\ifCLASSOPTIONtwocolumn
    \includegraphics[width=0.9\textwidth]{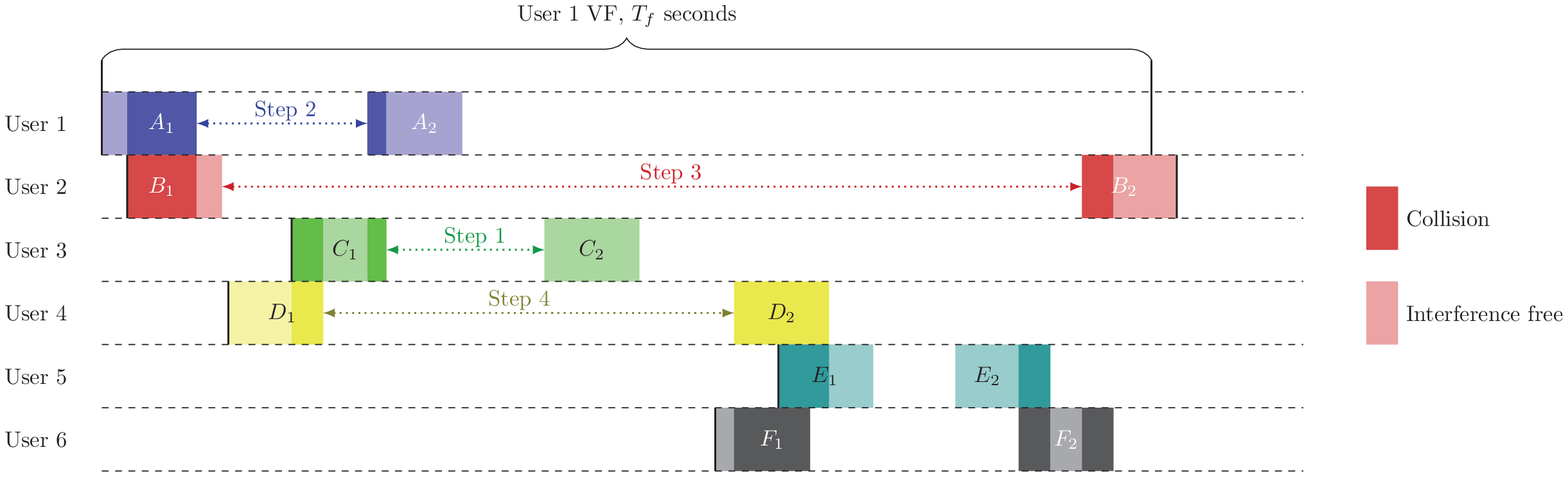}
\else
    \includegraphics[width=\textwidth]{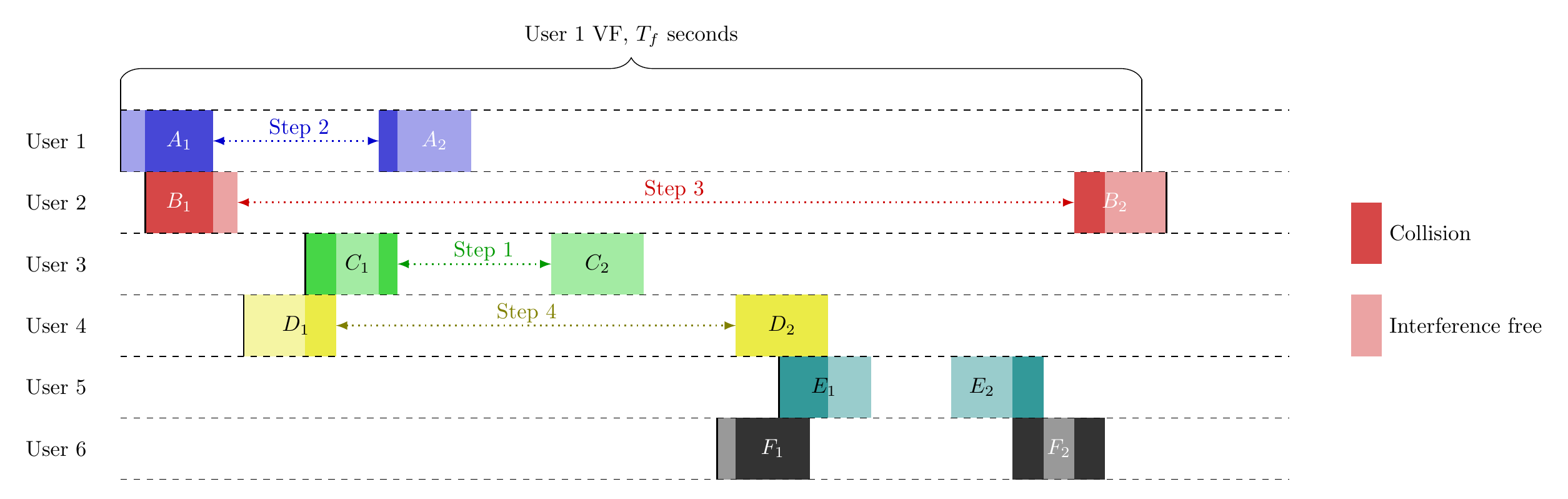}
\fi
\caption{\ac{SIC} procedure in \ac{ECRA}, first phase. The decoder starts looking for replicas that can be successfully decoded. The first to be found is replica $C_2$, which is collision-free. After successfully decoding, the information on the location of replica $C_1$ is retrieved from the header. So, the data carried by $C_2$ can be re-encoded, re-modulated, frequency offset and sampling epoch are superimposed on the signal, and its interference contribution is removed from both locations within the received signal. The interference caused on replica $A_2$ is now removed. The decoder can successfully decode also replica $A_2$, and applying the same procedure, remove its interference together with the one of replica $A_1$. Now, replica $B_1$ is collision-free, can be successfully decoded and its interference contribution together with the one of replica $B_2$ can be removed. Finally, replica $D_1$ also collision-free is correctly decoded, and removed from the received signal together with its twin $D_2$. Unfortunately user $5$ and $6$ replicas are in a collision pattern that cannot be resolved by \ac{SIC} only, and still remain in the received signal after the end of the first phase.}
\label{fig:CRA_SIC}
\end{figure*}
At the receiver, \ac{ECRA} follows a two phase procedure, in order to decode the received packets. The receiver will operate with a sliding window, similarly to \cite{Meloni2012, DeGaudenzi2014}. The decoder starts operating on the first $\Wind$ samples, with $\Wind$ the designed window size.

\subsubsection{\ac{SIC} phase}
During the first phase, the decoder seeks for replicas that can be successfully decoded. Making the use of the example shown in Fig. \ref{fig:CRA_SIC} where a degree $\dg=2$ has been selected, we describe the \ac{SIC} procedure. The first replica that can be decoded is $C_2$, assuming that very limited interference can be counteracted by the error correcting code. Thanks to the pointer to the position of all replicas of this user in the header, the decoder can retrieve the position of replica $C_1$ as well. In this way, replica $C_2$ can be re-encoded, re-modulated, frequency offset and sampling epoch are superimposed on the signal, and its interference contribution is removed from both locations within the received signal. In the following we assume ideal \ac{SIC}, i.e. the entire interference contribution is removed from the received signal. Replica $A_2$ is now released from the interference, and can also be correctly decoded. In this scenario, the \ac{SIC} procedure is iterated until none of the replicas can be successfully decoded anymore. At the end of \ac{SIC}, users $1, 2, 3, 4$ can be correctly decoded, while users $5$ and $6$ remain still unresolved, due to the presence of reciprocal interference that cannot be counteracted by the channel code.\footnote{In this example, we assumed a channel code able to counteract only very limited interference. In general, the code rate can be lowered to support higher levels of interference, thus possibly, leading to successful decoding. Unfortunately, there are still interference patterns in which the channel code alone is not able to counteract the interference, and prevents \ac{SIC} to resolve the collision.}

\subsubsection{Combining phase}
\acreset{SC,EGC,MRC,ECRA-SC,ECRA-MRC}
In the second phase of \ac{ECRA}, combining techniques are applied on the received packets unable to be decoded in first phase, and on these \emph{combined observations} decoding is attempted. The formal definition of a combined observation is as follows:
\begin{definition}[Combined observation]
\begin{sloppypar}
Consider the $\dg$ observations of the $\user$-th packet, $\rxVec^{(\user,1)}, \rxVec^{(\user,2)}, ..., \rxVec^{(\user,\dg)}$ with $\rxVec^{(\user,\replica)}=\left(\rx_0^{(\user,\replica)}, \rx_1^{(\user,\replica)}, ..., \rx_{\numSym-1}^{(\user,\replica)}\right)$. We define the \emph{combined observation} the vector
\begin{equation}
\rxVec^{(\user)}=\left(\rx_0^{(\user)}, \rx_1^{(\user)}, ..., \rx_{\numSym-1}^{(\user)}\right)
\end{equation}
with $\rx_i^{(\user)}$ being a suitable function of the individual observation samples $\rx_i^{(\user,1)},\rx_i^{(\user,2)},...,\rx_i^{(\user,\dg)}$, i.e.
\begin{equation}
\rx_i^{(\user)}:=f \left(\rx_i^{(\user,1)},\rx_i^{(\user,2)},...,\rx_i^{(\user,\dg)}\right).
\end{equation}
%A packet composed by symbols belonging to at least two replicas of the same user is denoted as \emph{combined packet}. Combined packets can be composed by sets of symbols belonging to different replicas or can have symbols being a weighted sum of the symbols coming from different replicas.
\end{sloppypar}
\end{definition}
Any of \ac{SC}, \ac{EGC} or \ac{MRC} \cite{Brennan1959,Jakes1974} can be applied in the second phase of \ac{ECRA}, although our focus will be on \ac{SC} and \ac{MRC}. If \ac{SC} is adopted, the combined observation is composed by the replica sections with the highest \ac{SINR}, i.e. for each observed symbol, the selection combiner chooses the replica with the highest \ac{SINR}. Hence, the instantaneous mutual information of the $\user$-th user combined observation, $i$-th symbol after \ac{SC} is
\begin{equation}
\mutInf \left(\sinr_{i}^{\mathrm{S}}\right) = \log_2\left( 1 + \max_\replica \left[\sinr_{i}^{(\user, \replica)}\right] \right).
\end{equation}
\begin{figure}
\centering
\ifCLASSOPTIONtwocolumn
    \includegraphics[width=\columnwidth]{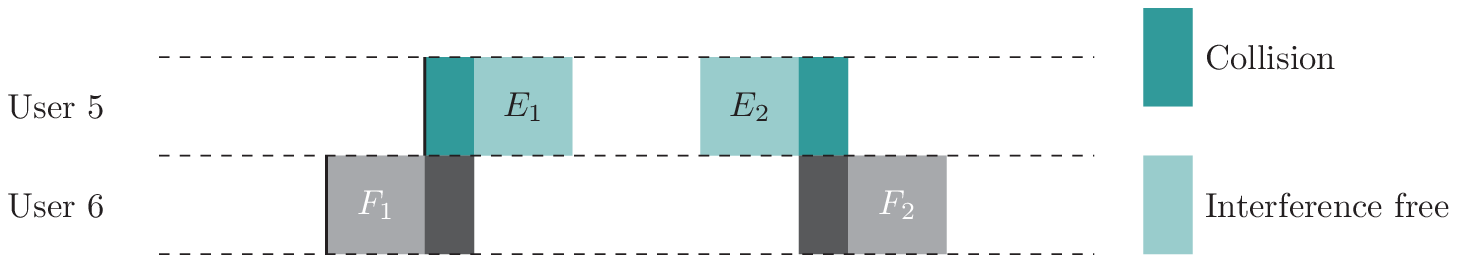}
\else%
    \includegraphics[width=0.6\columnwidth]{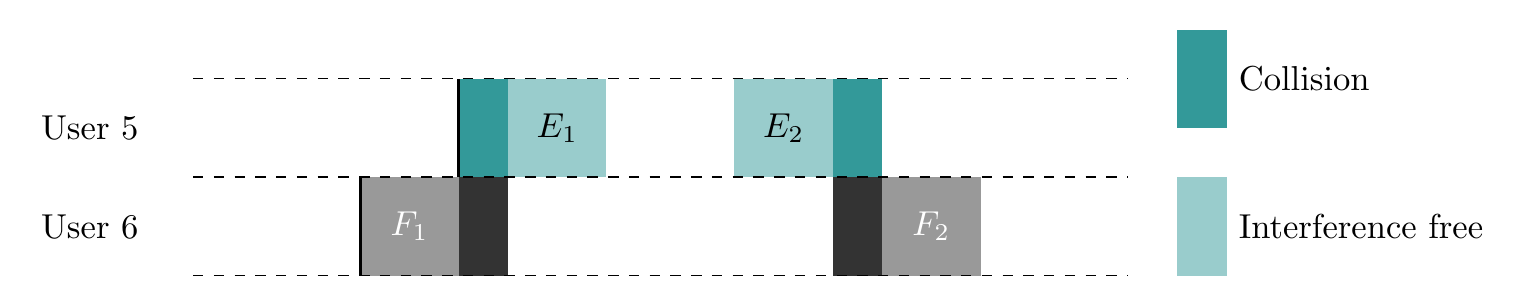}
\fi
\caption{Example of a collision pattern blocking \ac{SIC}. Different portions of replicas $F_1$ and $F_2$ are collision-free. When \ac{SC} is applied, \ac{ECRA} selects these portions, creates a combined observation and attempts decoding on it.}
\label{fig:CRA_SIC_2}
\end{figure}
Fig. \ref{fig:CRA_SIC_2} depicts the situation at the beginning of the second phase, for the example presented in Fig. \ref{fig:CRA_SIC}. The selection combiner selects the first part of replica $F_1$, and the second part of $F_2$, building a combined observation free from interference.

In \ac{ECRA-MRC} instead, each replica observed symbol of a given user is weighted proportional to its root mean squared received signal level \cite{Brennan1959}. In this way, the \ac{SINR} at the output of the combiner is the sum of all replicas \ac{SINR}s. It is also known from literature that, \ac{MRC} is optimal if the interference on each replica is independent \cite{Winters1984}. The instantaneous mutual information of the $\user$-th user combined observation $i$-th symbol after \ac{MRC} is
\begin{equation}
\mutInf \left(\sinr_{i}^{\mathrm{M}}\right) = \log_2\left(1+\sum_{\replica=1}^{\dg} \sinr_{i}^{(\user, \replica)}\right).
\end{equation}
The decoder outcome after \ac{SC} or \ac{MRC} is modeled substituting the expression of $\mutInf(\sinr_{i})$ with $\mutInf \left(\sinr_{i}^{\mathrm{S}}\right)$ or $\mutInf \left(\sinr_{i}^{\mathrm{M}}\right)$ in equation \eqref{eq:mi_dec}, and adopting the same condition as in equation \eqref{eq:dec_res}. When decoding is successful, the packet is re-encoded, re-modulated and its interference contribution is removed in all the positions within the frame, where the replicas of the decoded user are placed. Combining and \ac{SIC} are iterated until, either all users are correctly decoded, or no more packets are present in the receiver window $\Wind$. The receiver window is then shifted forward by $\WindSh$ samples, and the procedure starts again.

\subsection{Summary and Comments}

The second step of \ac{ECRA} needs complete knowledge of the replicas position of the remaining users in the frame. Although stringent, this requirement can be addressed in two practical ways: either adopting dedicated pointers to the replicas locations in the header, or exploiting correlation techniques for detection and combining of the replicas, prior to decoding. The former solution adopts a pseudo-random seed, that is used at the receiver, for retrieving the information on all replicas position of the decoded user. This option was proposed first in \cite{Casini2007}, for slotted protocols, but can be extended also to \ac{ECRA}. The pseudo-random seed is used for generating the relative time offset between replicas, and together with the replica sequence number, allows to identify the replicas locations. In \cite{Clazzer2013} it is shown that, in the low to moderate channel traffic regions, low probability of interference in the header can be found. In the high channel traffic region instead, replicating the header twice is beneficial. Moreover, if a dedicated channel code is introduced for protecting the header, lower header loss probability are expected.\footnote{Dedicated channel code applied to the headers, can allow retrieving the information about replica locations, although the packet itself is cannot be decoded due to collisions.}

When correlation techniques are adopted, no overhead due to a dedicated field in the header is necessary, and replicas are detected and combined before decoding \cite{Clazzer2017_Corr}. A two-phase procedure is proposed in \cite{Clazzer2017_Corr}: first, transmitted replicas are detected, and second, replicas belonging to the same user are matched. Specifically, exploiting the presence of preamble at the start of a replica, common to every user and replica, a non-coherent soft-correlation metric is adopted for detection. The correlator needs to distinguish between two hypothesis, the presence of a preamble, or simply the presence of noise and, potentially, interference. A threshold test is exploited, and the correlator outputs a decision for each sample of the recorded signal. Whenever the correlator exceeds a defined threshold, the sample is declared as start of a candidate replica. In the second phase, for each candidate replica, the one belonging to the same user are sought, within the set of candidate replicas. Similarly to the first phase, a non-coherent soft-correlation metric is a adopted. On the other hand, in the second phase, we can exploit the information along the whole packet, since replicas of the same user are identical. This helps the performance to improve enormously. In fact, we can use correlation over thousands of symbols, instead of only few tens as per the preamble. In the mentioned work, we proposed also two enhancements to the technique. In the first place, if users transmit their replicas within discrete intervals, the second phase can benefit of a reduction in complexity. Candidate replicas sent out of the discrete intervals can be discarded a-priori, without requiring the correlation to take place. Furthermore, if the interference power is estimated, the first phase can be enhanced. Starting from the approximate likelihood ratio test, we developed an interference-aware metric. In the paper, results on both first and second phase are presented, showing very good performance. The comparison between ideal detection and matching compared to the two-phase approach shows a very limited degradation in spectral efficiency.

% First the detection of the replicas on the channel is carried out and second, the matching between replicas belonging to the same user is performed. In both phases a simple non-coherent soft-correlation metric is applied. In the first phase, the sync word common to all replicas is sought in the received signal. All positions exceeding the defined threshold are marked as candidate replicas. In the \RTwo{second} phase, for each of the candidate replicas identified in the first phase, its signal is correlated with all other candidates. In \RTwo{this phase,} the correlation is data-aided so that the entire replica signal including the data part is employed in the non-coherent soft-correlation. The candidate replica showing the highest correlation is combined with the considered one and decoding is attempted. The channel decoder is assumed to be able to detect unresolvable errors with high probability. In order to reduce the number of possible combinations in the \RTwo{second} phase, the replicas are allowed to be sent only in a finite number of discrete positions within their \ac{VF}. In this way, only a subset of candidates replicas are possible twins of the considered one. In \cite{Clazzer2017_Corr} we have shown that using this approach very limited performance degradation are expected compared to the ideal assumption of perfect knowledge about the replicas position.

A similar correlation approach have been proposed by the authors of \cite{Bui2015}, but in a time-synchronous scenario. Two main differences can be identified, firstly the technique proposed for the asynchronous scenario adopts a two-phase non-coherent correlation approach where, in the first phase only, the detection is performed. Secondly, a discretization of the time instants in which a user can transmit its replicas is proposed, so to reduce the number of correlations to be performed. In other words, not all combinations of candidate replicas are allowed.

\ac{MRC} combining technique requires the knowledge of the \ac{SINR} symbol-by-symbol, in order to choose the optimal weights \cite{Brennan1959} beforehand the combination is done. In case this information cannot be retrieved, combining can be applied with equal weights for all the symbols, i.e. \ac{EGC}.

The scenarios under consideration in the work of \cite{ZigZag}, and its extension \cite{SigSag}, are similar to the one that can block the \ac{SIC} procedure (see Fig. \ref{fig:CRA_SIC_2}), although some differences in the solutions between their work and \ac{ECRA} can be identified. \ac{ECRA} creates the combined observation and tries decoding on it, while \cite{ZigZag} requires an iterative demodulation procedure within packet portions, that may increases the overall packet decoding delay. Furthermore, in \cite{ZigZag} an error in one decoded bit propagates to the entire packet unless compensated by further errors. This is due to the iterative procedure applied, which subtracts the uncorrect bit from the same packet in the second collision, while in \ac{ECRA} an error in one decoded bit will not affect any other portion of the packet. 
\section{Packet Loss Rate Analysis at Low Channel Load}
\label{sec:PLR}
\begin{figure*}[t]
\ifCLASSOPTIONtwocolumn
    \begin{subfigure}{2\columnwidth}
\else%
    \begin{subfigure}{\columnwidth}
\fi
\centering\captionsetup{width=.9\textwidth}
\includegraphics[width=\columnwidth]{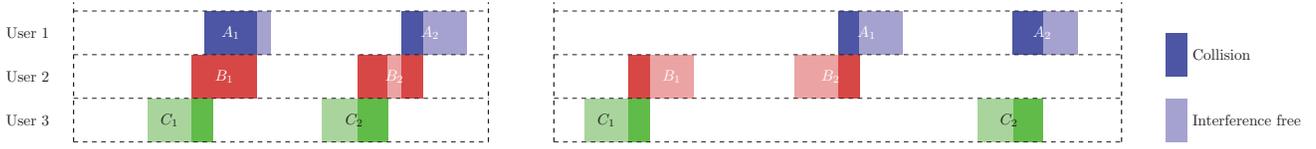}
   \caption{Two types of collision clusters with $|\CollClus|=3$ and with $\dg=2$. If the error correcting code~$\Code$ is not able to counteract the interference of any replica, the collision clusters are also $\Code$-unresolvable collision patterns.}
   \label{fig:cc_3}
\end{subfigure}
\begin{center}
\ifCLASSOPTIONtwocolumn
    \begin{subfigure}{0.8\columnwidth}
\else%
    \begin{subfigure}{0.5\columnwidth}
\fi
\centering\captionsetup{width=.9\columnwidth}
\includegraphics[width=\columnwidth]{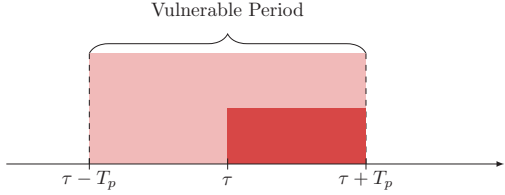}
   \caption{Vulnerable period for a packet transmitted with an asynchronous \ac{RA} protocol, under the collision channel model.}
   \label{fig:vul_per1}
\end{subfigure}
\ifCLASSOPTIONtwocolumn
    \hspace{1.5cm}
\fi
\ifCLASSOPTIONtwocolumn
    \begin{subfigure}{0.8\columnwidth}
\else%
    \begin{subfigure}{0.5\columnwidth}
\fi
\centering\captionsetup{width=.9\columnwidth}
\includegraphics[width=\columnwidth]{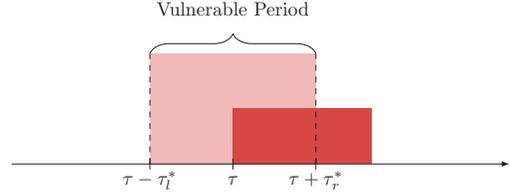}
   \caption{Vulnerable period for a packet transmitted with an asynchronous \ac{RA} protocol, with an error correcting code~$\Code$.}
   \label{fig:vul_per2}
\end{subfigure}
\end{center}
\caption{Examples of collision clusters, $\Code$-unresolvable collision patterns and vulnerable periods.}
\label{fig:cc_and_vul_per}
\end{figure*}

In this Section a \ac{PLR} approximation, tight for low channel load conditions, is derived. Packet losses are caused by particular interference patterns that \ac{SIC} is not able to resolve. In the slot synchronous \ac{RA} protocols, these patterns are analogous to the stopping sets present in \ac{LDPC} codes \cite{Ivanonv_2015_Letter}, and can be analyzed exploiting tools from coding theory, and graph theory. In the asynchronous \ac{RA} schemes, a graph representation is not straightforward, since no discrete objects as slots are present anymore. Therefore, we resort to investigate the collision patterns that involve two users only, with a generic degree $\dg$, and conjecture that these are the patterns driving the \ac{PLR}, especially at low channel loads. In the next section, the approximation of the \ac{PLR} is compared with Monte Carlo simulations, in order to verify its tightness. A set of definitions are required for the analysis.
\begin{definition}[Collision cluster $\CollClus$]
Consider a subset $\CollClus$ of users. Assume that packets of all users in $\CollClus^{c}$ (complementary of the subset $\CollClus$) have been successfully decoded. The subset $\CollClus$ is referred to as \emph{collision cluster} iff no packet replicas for the users in $\CollClus$ is collision-free.
\end{definition}
Under the assumption of collision channel, none of the users in the collision cluster can be successfully decoded. Conversely, when a channel code $\Code$ is employed by each transmitted packet, the collision cluster might be resolvable, leading to the following definition.
\begin{definition}[$\Code$-unresolvable collision pattern]
\acreset{UCP}
Given each packet encoded with a channel code $\Code$, a \emph{\ac{UCP}} $\UCP$ is a collision cluster where no user in the set can be successfully decoded.
\end{definition}
Every \ac{UCP} is also a collision cluster, but not viceversa. In order to evaluate the probability of \ac{UCP} involving two users only, a generalization of the definition of vulnerable period \cite{Kleinrock1976_book} is required.
\begin{definition}[$\Code$-vulnerable period for $|\CollClus|=2$]
Consider the transmission of a packet protected with a channel code $\Code$ between time $\RStart$ and $\RStart + \pkLen$. The packet's \emph{$\Code$-vulnerable period} is the interval of time $\left[\RStart - \VL, \RStart + \VR\right]$ in which the presence of a single interferer leads to a failure in the decoding.
\end{definition}
Hence, the vulnerable period duration $\Vpd$ is defined as
\begin{equation}
\Vpd = \VL + \VR.
\end{equation}
\begin{sloppypar}
In slotted synchronous schemes under the collision channel model, $\VL=0$ and $\VR = \pkLen$, so $\Vpd = \pkLen$. For asynchronous schemes in general and therefore for \ac{ECRA}, it holds ${\VL = \VR \triangleq \Vg}$. The vulnerable period duration for asynchronous schemes is $\Vpd = 2 \Vg$. Considering the collision channel model, the vulnerable period duration is then $\Vpd = 2 \pkLen$. So, the duration of packets' vulnerable period is doubled in asynchronous schemes w.r.t. comparable synchronous ones \cite{Kleinrock1976_book}. Examples of collision clusters, $\Code$-unresolvable collision patterns and vulnerable periods are provided in Fig. \ref{fig:cc_and_vul_per}.
\end{sloppypar}
%\begin{figure}
%\centering
%\subfigure[\RTwo{Two types of collision clusters with $|\CollClus|=3$ and with $\dg=2$. If the error correcting code~$\Code$ is not able to counteract the interference of any replica, the collision clusters are also $\Code$-unresolvable collision patterns.}]{
%    \includegraphics[width=\columnwidth]{./Figures_TikZ/cc_3}
%    \label{fig:cc_3}
%}\\
%\subfigure[\RTwo{Vulnerable period for a packet transmitted with an asynchronous \ac{RA} protocol, under the collision channel model.}]{
%    \includegraphics[width=0.47\columnwidth]{./Figures_TikZ/vul_period1}
%    \label{fig:vul_per1}
%}
%\subfigure[\RTwo{Vulnerable period for a packet transmitted with an asynchronous \ac{RA} protocol, with an error correcting code~$\Code$.}]{
%    \includegraphics[width=0.47\columnwidth]{./Figures_TikZ/vul_period2}
%    \label{fig:vul_per2}
%}
%\caption{\RTwo{Examples of collision clusters, $\Code$-unresolvable collision patterns and vulnerable periods.}}
%\label{fig:cc_and_vul_per}
%\end{figure}

\subsection{Packet Loss Rate Approximation}
\label{sec:Approx_PER}
In this section, we derive an approximation of the \ac{PLR}, denoted as $\plr$. The approach follows \cite{Sandgren2017}, extending the investigation to asynchronous schemes. Let us consider the user $\user$. We denote with $\AllUCP$ the set of all possible \ac{UCP} that cause the loss of user $\user$ packets, with $\UCP$ the index of the \ac{UCP} set considered when writing the union bound, and with $\UCPcons$ the unique type of \ac{UCP} that we assume to drive the \ac{PLR} performance $\plr$. Let $\fraTp = \fraLen/\pkLen$ denote the \ac{VF} length measured in packet durations, and $\nVp=\lfloor \fraLen/\Vpd \rfloor$ denote the number of disjoint vulnerable periods per \ac{VF}.\footnote{It is possible that, the vulnerable period of the replicas of one user overlap. This is the case when the relative delay between replicas is smaller than the vulnerable period. On the other hand, this probability is reasonably small for virtual frames of interest.} Clearly $\fraTp \geq \dg$. The \ac{PLR} can be approximated with
\begin{equation}
\begin{split}
\label{eq:plr_approx}
\plr &= \Pr\left\{\bigcup_{\UCP \in \AllUCP} \user \in \UCP \right\} \leq \sum_{\UCP \in \AllUCP} \Pr\left\{\user \in \UCP \right\} \approx \Pr\left\{\user \in \UCPcons \right\} \\
&= \sum_{m=2}^{\infty} \frac{e^{-\fraTp \load}\left( \fraTp \load \right)^m}{m!}\Pr\left\{ \user \in \UCPcons | m \right\}.
\end{split}
\end{equation}
The probability $\plr$ is first bounded from above with the union bound and then approximated considering only one type of \ac{UCP}, i.e. $\UCPcons$. Finally, we take the expectation of the number of active users over the $\user$-th user \ac{VF}. The \ac{UCP} $\UCPcons$ considered in the analysis is formed by two users only with a generic degree $\dg$. The probability that, the considered user $\user$ belongs to the \ac{UCP} $\UCPcons$ formed by two users, is approximated as
\begin{equation}
\label{eq:prob_UCP}
\Pr\left\{ \user \in \UCPcons | m \right\} \approx \frac{\mU(\UCPcons,m) \sL(\UCPcons)} {\UL(\UCPcons)} \frac{2}{m}.
\end{equation}
\begin{sloppypar}
Where, we denote with $\mU(\UCPcons,m)$ the number of possible combinations of $m$ users, taken two by two, i.e. ${\mU(\UCPcons,m) = \binom{m}{2}}$. The second term $\sL(\UCPcons)$, accounts for the number possible placements of $\dg$ replicas. Since the first replica of user $\user$ is sent immediately, only the remaining $\dg-1$ are free to be sent within a delay selected uniformly at random, and not exceeding a \ac{VF} duration. In terms of vulnerable periods, the number of possible placements of the remaining $\dg-1$ replicas is $\sL(\UCPcons) \approx \binom{\nVp-1}{\dg-1}$. Similarly, the number of ways in which two users can select their position for the replicas, follows $\UL(\UCPcons) \approx \nVp \binom{\nVp-1}{\dg-1}^2$. The probability that the two users (including user $\user$) placed all their $\dg$ replicas in the reciprocal vulnerable periods is $1/\UL(\UCPcons)$. Finally, the probability that the two users belong to the \ac{UCP} $\UCPcons$ is $2/m$. Substituting into equation \eqref{eq:prob_UCP} we get
\end{sloppypar}
\begin{equation}
\label{eq:prob_UCP2}
\Pr\left\{ \user \in \UCPcons | m \right\} \approx \frac{\binom{m}{2}}{\nVp \binom{\nVp-1}{\dg-1}} \frac{2}{m} = \frac{\binom{m}{2}}{\dg \binom{\nVp}{\dg}} \frac{2}{m}.
\end{equation}
Finally, inserting in equation \eqref{eq:plr_approx} the result of equation \eqref{eq:prob_UCP2} we can approximate the \ac{PLR} $ \plr$ as
\begin{equation}
\label{eq:plr_approx_final}
\plr \approx \sum_{m=2}^{\infty} \frac{e^{-\fraTp \load}\left( \fraTp \load \right)^m}{m!} \frac{\binom{m}{2}}{\dg \binom{\nVp}{\dg}} \frac{2}{m}.
\end{equation}

The \ac{PLR} approximation directly depends on the vulnerable period duration, via $\nVp$. In the next Sections, the vulnerable period duration is computed for two scenarios of interest, including the \ac{MRC} case.

\subsection{Vulnerable Period Duration for Asynchronous \ac{RA} with \ac{FEC}}

\begin{sloppypar}
In this scenario, packets, or replicas for schemes adopting them, are protected by a channel code so that not all collisions are destructive. The only \ac{UCP} to be considered is the one involving two users and their packets or replicas. We recall that perfect power control is assumed so that both users are received with the same power $\usPw$. Without loss of generality, we focus on a specific packet, or replica, involved in an \ac{UCP} of type $\UCPcons$, which has a first section free of interference and a second part interfered. The selected rate $\rate$ determines what it is the minimum fraction of interference-free packet, or replica, $\intFreeAsy$ that still allows correct decoding, i.e.
\begin{equation}
\label{eq:app_a_3}
\intFreeAsy \log_2\left(1 + \frac{\usPw}{\noisePw}\right)+( 1 - \intFreeAsy )\log_2\left(1 + \frac{\usPw}{\noisePw + \usPw}\right) = \rate.
\end{equation}
For the sake of simplicity we denote with
\begin{equation}
\begin{split}
\rateFree&= \log_2\left(1 + \frac{\usPw}{\noisePw}\right)\\
\rateInt&= \log_2\left(1 + \frac{\usPw}{\noisePw + \usPw}\right)
\end{split}
\end{equation}
and we solve equation \eqref{eq:app_a_3} for $\intFreeAsy$
\begin{equation}
\label{eq:app_a_4}
\intFreeAsy = \frac{\rate-\rateInt}{\rateFree-\rateInt}.
\end{equation}
Equation \eqref{eq:app_a_4} is valid for $\rate \geq \log_2\left(1 + \frac{\usPw}{\noisePw + \usPw}\right)$. In fact, for $\rate < \rateInt$, no \ac{UCP}s involving only two users can be observed, and regardless the level of interference, packets involved in collisions with only one other packet can be always decoded. In this way,
\begin{equation}
\label{eq:app_a_5}
\intFreeAsy = \left\{
\begin{array}{rl}
\frac{\rate - \rateInt}{\rateFree - \rateInt} & \mbox{for $\rate \geq \rateInt$}\\
0 & \mbox{for $\rate < \rateInt$}
\end{array}
\right .
\end{equation}
It is worth noticing that $\intFreeAsy$ is constrained to $0\leq \intFreeAsy \leq1$, since the selectable rate $\rate$ is ${\rate \leq \log_2\left(1 + \frac{\usPw}{\noisePw}\right) = \rateFree}$ for reliable communication.
\end{sloppypar}

In this way, $\Vg = \intFreeAsy \pkLen$ and therefore the vulnerable period is reduced to $\Vpd = 2 \Vg = 2 \intFreeAsy \pkLen$. And finally $\nVp=\lfloor \fraLen/\Vpd \rfloor=\lfloor \fraLen/(2 \intFreeAsy \pkLen) \rfloor$. Inserting the value of $\nVp$ in equation \eqref{eq:plr_approx_final} gives the final expression of the \ac{PLR} approximation for asynchronous \ac{RA} schemes using replicas. Note that for $\intFreeAsy \rightarrow 0$, $\nVp \rightarrow + \infty$ and therefore the \ac{PLR} approximation in equation \eqref{eq:plr_approx_final} tends to $0$.

\subsection{Vulnerable Period Duration for Asynchronous \ac{RA} with \ac{MRC} and $\dg=2$}
\label{sec:low_bound_ECRA_MRC}

Similarly to the previous section, $\UCPcons$ is the considered \ac{UCP} where two users are interfering each other and they are received with the same power $\usPw$. In this scenario the degree is fixed to $\dg=2$. Focus is on the combined observation, after \ac{MRC}. Without loss of generality, it is assumed that the first section of both replicas is free of interference, while there is a second part where just one replica is interfered and finally there is the last part where both replicas are interfered. We aim at computing the minimum combined observation portion interference free $\intFreeMRC$, that is required for correctly decoding the user after \ac{MRC}. It holds
\ifCLASSOPTIONtwocolumn
    \begin{equation}
    \begin{aligned}
    \label{eq:mrc1}
    &\intFreeMRC \log_2\left(1+2 \frac{\usPw}{\noisePw}\right) + \intOne \log_2\left(1+\frac{\usPw}{\noisePw} + \frac{\usPw}{\noisePw + \usPw}\right)\\ &+ (1-\intFreeMRC -\intOne )\log_2\left(1+ 2 \frac{\usPw}{\noisePw + \usPw}\right) = \rate.
    \end{aligned}
    \end{equation}
\else
    \begin{equation}
    \label{eq:mrc1}
    \intFreeMRC \log_2\left(1+2 \frac{\usPw}{\noisePw}\right) + \intOne \log_2\left(1+\frac{\usPw}{\noisePw} + \frac{\usPw}{\noisePw + \usPw}\right) + (1-\intFreeMRC -\intOne )\log_2\left(1+ 2 \frac{\usPw}{\noisePw + \usPw}\right) = \rate.
    \end{equation}
\fi
For the sake of simplicity, we denote with
\begin{equation}
\begin{split}
\rateFree &= \log_2\left(1+2 \frac{\usPw}{\noisePw}\right)\\
\rateIntO &= \log_2\left(1+\frac{\usPw}{\noisePw} + \frac{\usPw}{\noisePw + \usPw}\right)\\
\rateIntT &= \log_2\left(1+ 2 \frac{\usPw}{\noisePw + \usPw}\right).
\end{split}
\end{equation}
So that equation \eqref{eq:mrc1} becomes
\begin{equation}
\label{eq:app_a_6}
\intFreeMRC \rateFree + \intOne \rateIntO + (1 - \intFreeMRC - \intOne ) \rateIntT = \rate.
\end{equation}

In order to solve equation \eqref{eq:app_a_6}, $\intOne$ is expressed as a function of $\intFreeMRC$, as $\intOne = \alpha \intFreeMRC$, where $0\leq\alpha\leq (1-\intFreeMRC)/\intFreeMRC$. When $\alpha=0$, there are no portions where only one out of the two replicas is interfered, while $\alpha=(1-\intFreeMRC)/\intFreeMRC$ represents the case when there are no portions where both replicas are interfered. Resolving \eqref{eq:app_a_6} for $\intFreeMRC$ gives
\begin{equation}
\label{eq:app_a_7}
\intFreeMRC = \frac{\rate-\rateIntT}{\rateFree-\rateIntT+\alpha(\rateIntO-\rateIntT)}.
\end{equation}
Also in this case, for $\rate<\rateIntT$, $\intFreeMRC=0$ which means that no \ac{UCP} involving two replicas can be found,
\begin{equation}
\label{eq:app_a_8}
\intFreeMRC= \left\{
\begin{array}{rl}
\frac{\rate-\rateIntT}{\rateFree-\rateIntT+\alpha(\rateIntO-\rateIntT)} & \mbox{for $\rate\geq \rateIntT$}\\
0 & \mbox{for $\rate<\rateIntT$}
\end{array}
\right .
\end{equation}
The average vulnerable period duration over the two replicas is $\Vpd = 2 \Vg = 2\left( \intFreeMRC + \frac{\intOne}{2} \right) \pkLen = 2\intFreeMRC\left(1 + \frac{\alpha}{2}\right) \pkLen$.\footnote{It is important to underline that, the expression of the average vulnerable period duration presented is valid no matter how the two replicas are interfered, i.e. also when the portions interfered are not both at the beginning of the packets.} And finally $\nVp=\lfloor \fraLen/\Vpd \rfloor=\lfloor \fraLen/\left(2\intFreeMRC\left(1 + \frac{\alpha}{2}\right) \pkLen \right) \rfloor$.

The presented analysis can be extended also to a higher number of replicas, i.e. $\dg>2$. The main difference will be in the number of packet sections that shall be taken into account, which corresponds to $\dg+1$, in general. The other key difference is in the number of parameters, which grows as $\dg$. 
\section{Performance Analysis}
\label{sec:simulations_all}

In this section, \acs{ECRA-SC} and \ac{ECRA-MRC} are compared with the reference \ac{CRA} protocol, as well as with ALOHA. For this first comparison two metrics are considered, the \ac{PLR} and the throughput. The throughput $\tp$ is defined as the expected number of successfully decoded packets per packet duration $\pkLen$,
\begin{equation}
\tp=(1-\plr)\, \load.
\end{equation}
The \ac{ECRA} algorithm is also compared against slot synchronous \ac{RA} protocols, as \ac{CRDSA}. Since a channel code $\Code$ is adopted in the proposed scheme, the throughput is not anymore a sufficient metric. %\ROne{Assuming that no power unbalance is present between the received packets, and the channel code cannot counteract any collision, } in slot synchronous protocols packets are decoded only if they are received collision-free. In the asynchronous case, instead, even without any power unbalance, a certain level of interference can be sustained and collisions can be resolved. The level of interference allowing correct decoding depends on the selected rate $\rate$. In the former case, regardless of the selected rate, the throughput performance remains the same, while in the latter lower rates lead to higher throughput. Nevertheless, lowering the rate decreases the information carried by each packet. This tradeoff is captured by the spectral efficiency $\se$,
In fact, thanks to the use of error correcting code at physical layer, even with equal received power, a certain level of interference can be sustained and collisions may be resolved. The level of interference allowing correct decoding depends on the selected rate $\rate$. Lowering the rate enables to resolve a higher number of collisions, but reduces the information carried by each packet. This tradeoff is captured by the spectral efficiency $\se$,
\begin{equation}
\label{eq:thr2}
\se=(1-\plr) \, \load \, \rate \qquad \mathrm{[bits/symbol]}.
\end{equation}
Although \ac{ECRA} can outperform considerably the ALOHA protocol, it entails a larger transmit energy per packet. In fact, this scheme assumes to replicate each packet sent in the frame $\dg$ times. In order to take into account the increase in energy per packet, we follow the approach of \cite{Abramson1977}, that was extended for slotted synchronous protocols as \ac{CRDSA} and \ac{IRSA} in \cite{Liva2011}. The \emph{normalized capacity} $\normCap$ is defined as the ratio between the maximum achievable spectral efficiency of one of the examined \ac{RA} scheme and the channel capacity of multiple access Gaussian channel under the same average power constraint. The average power constraint takes into account, the channel load and the number of replicas, so to directly reflect variations in the energy per packet. The idea is to compute the maximum spectral efficiency of the asynchronous \ac{MAC} schemes (\ac{ECRA-SC} or \ac{ECRA-MRC}) and normalize it to the sum rate capacity of the multiple access Gaussian channel $\refCap =\log_2(1+\PwAg/\noisePw)$. This is done fixing the average aggregate received signal power $\PwAg$ equal in all the schemes. In this way, for the \ac{RA} protocols the user transmission power $\usPwTx$ takes into account the fact that the channel is used intermittently but $\dg$ times w.r.t. ALOHA, i.e. $\usPwTx=\frac{\PwAg}{\load\cdot \dg}$. The ultimate performance of the asynchronous \ac{RA} schemes is given by the maximum spectral efficiency $\maxSe$ defined as
\begin{equation}
\label{eq:max_se}
\maxSe=\max_{\rate\in \left[0,..,\maxRate\right]}\tp(\load) \, \rate
\end{equation}
\begin{sloppypar}
where for each channel traffic value, the rate $\rate$ which maximizes the spectral efficiency is chosen.\footnote{The maximum rate for reliable communication $\maxRate$ is ${\maxRate=\log_2(1+\usPwTx/\noisePw)}$ and depends upon the selected channel load $\load$.} Unfortunately, the throughput expression $\tp(\load)$ is not available in closed form for \ac{ECRA-SC} and \ac{ECRA-MRC}, so only a numerical evaluation of equation \eqref{eq:max_se} is possible. The normalized capacity $\normCap$ is defined as
\end{sloppypar}
\begin{equation}
\label{eq:efficiency}
\normCap = \frac{\maxSe}{\refCap},
\end{equation}
where, depending on the \ac{RA}, a different expression of $\maxSe$ will be used.

\subsection{Numerical Results}
\label{sec:simulations}

\begin{sloppypar}
In the following, numerical results for \ac{ECRA-SC} and \ac{ECRA-MRC} schemes are presented.
%In the set of simulations, the symbol duration $\symLen$ is normalized to one second, i.e. $\symLen=1$ [s].
The packets sent by the users are composed by $\numBit=1000$ bits, which translate into $\numSym=\left(\numBit/\rate\right)$ symbols. The transmission period is then ${\pkLen = \symLen\,\numSym}$. The \ac{VF} duration $\fraLen$ is selected to be equal to $200$ packet durations, i.e. $\fraLen=200\,\pkLen$. We recall that, the number of users generating traffic follows a Poisson distribution, with mean $\load$ measured in packets per $\pkLen$ durations, and each of the users transmits $\dg=2$ replicas per generated packet. The decoder operates on a window of $\Wind = 3\,\fraLen = 600\,\pkLen$ and once either the maximum number of \ac{SIC} iterations is expired or no more packets can be successfully decoded, it is shifted forward by $\WindSh = 20\,\pkLen$. Ideal interference cancellation is assumed and the block interference model introduced in Section \ref{sec:int_model} is used for determining the successful decoding of a packet. Since the physical layer is abstracted, no frequency offset or phase offset are considered in the presented simulation results.\footnote{The impact of phase noise is not taken into account in the reported numerical results, due to the physical layer abstraction. The effect of phase noise is non-trivial and is therefore, left as part of future work.}
\end{sloppypar}

\begin{figure}
\centering
\ifCLASSOPTIONtwocolumn
    \includegraphics[width=0.95\columnwidth]{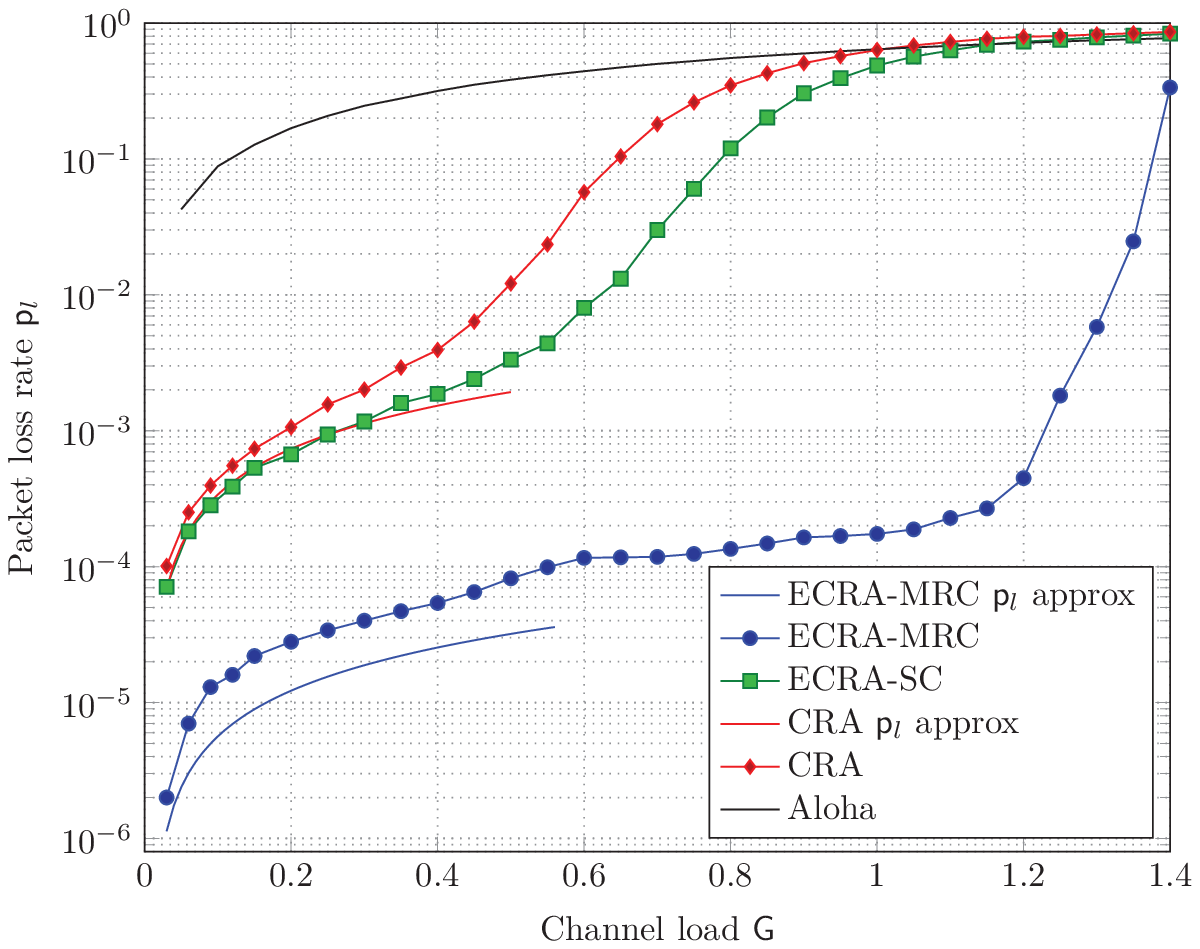}
\else%
    \includegraphics[width=0.7\columnwidth]{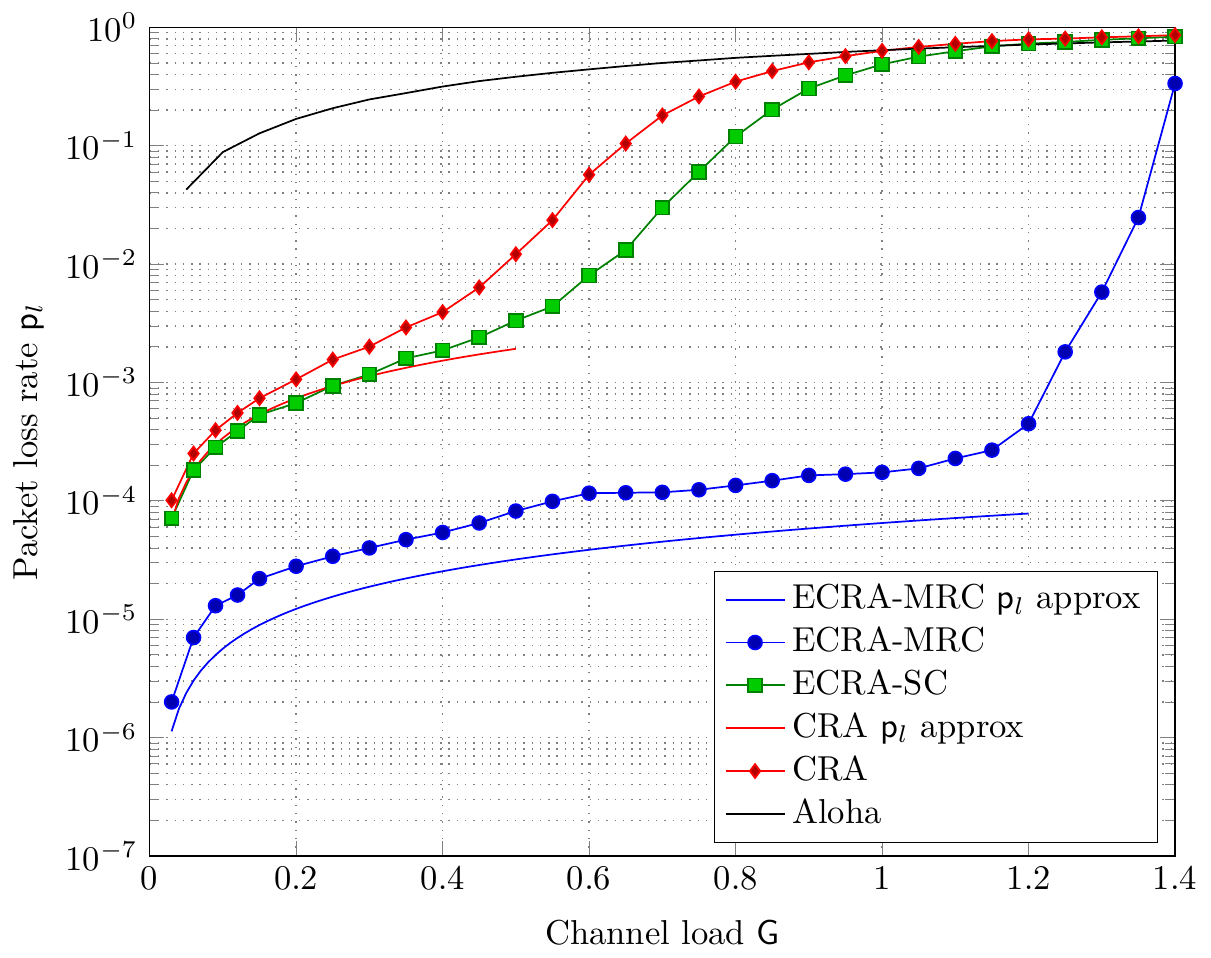}
\fi
\caption{Packet error rate $\plr$ vs. channel load $\load$ for ALOHA,
\ac{CRA}, \ac{ECRA-SC} and \ac{ECRA-MRC}, $\usPw/\noisePw=6$ dB and $\rate=1.5$.} %The average value of $\alpha$ used in the approximation of $\plr$ for \ac{ECRA-MRC} is derived through Monte Carlo simulations.
\label{fig:PER}
\end{figure}

We present first the simulations of the throughput and \ac{PLR} for both \ac{ECRA-SC} and \ac{ECRA-MRC}. For reference purposes also \ac{CRA} and the ALOHA protocols are depicted in the figures. The assumptions are $\usPw/\noisePw=6$ dB and $\rate=1.5$ equal for all users.
\begin{figure}
\centering
\ifCLASSOPTIONtwocolumn
    \includegraphics[width=0.95\columnwidth]{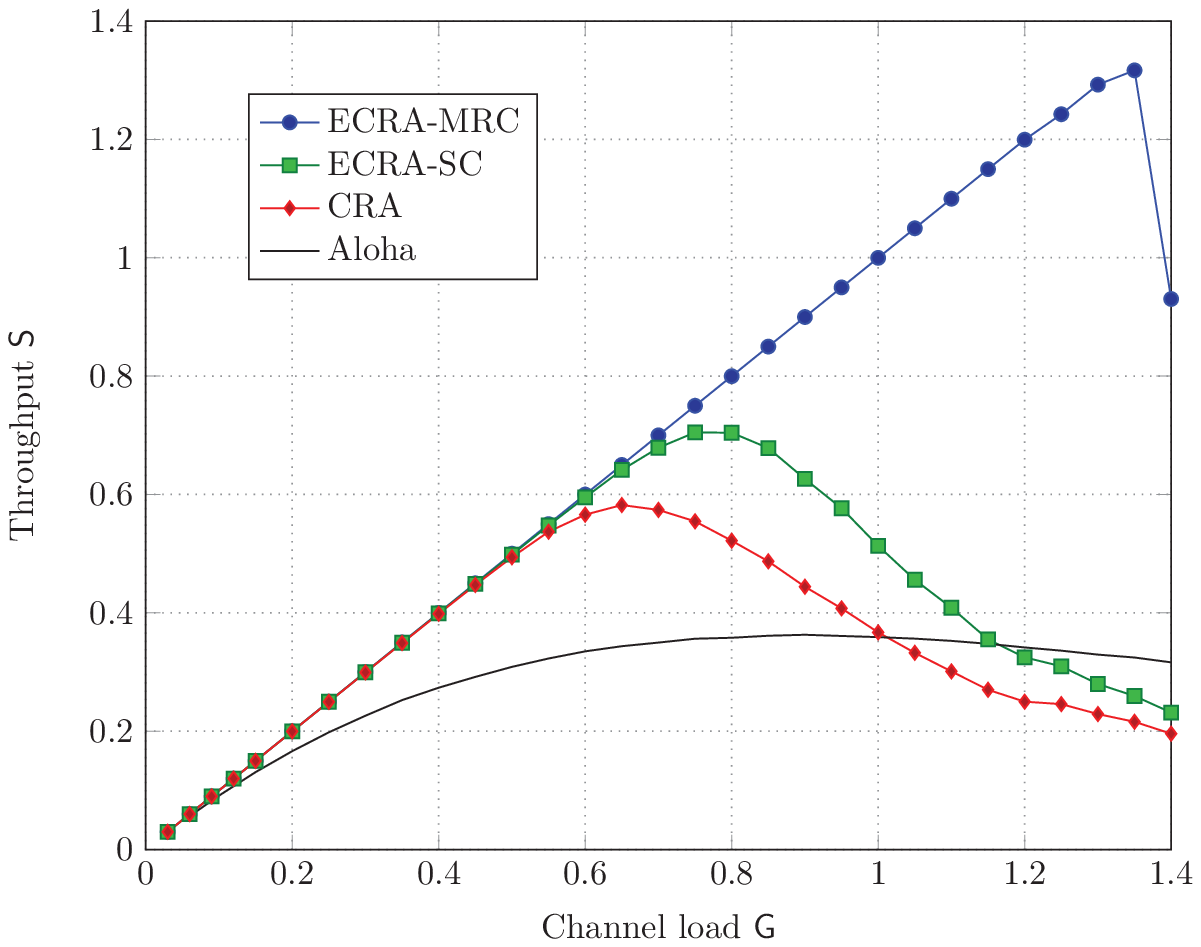}
\else%
    \includegraphics[width=0.7\columnwidth]{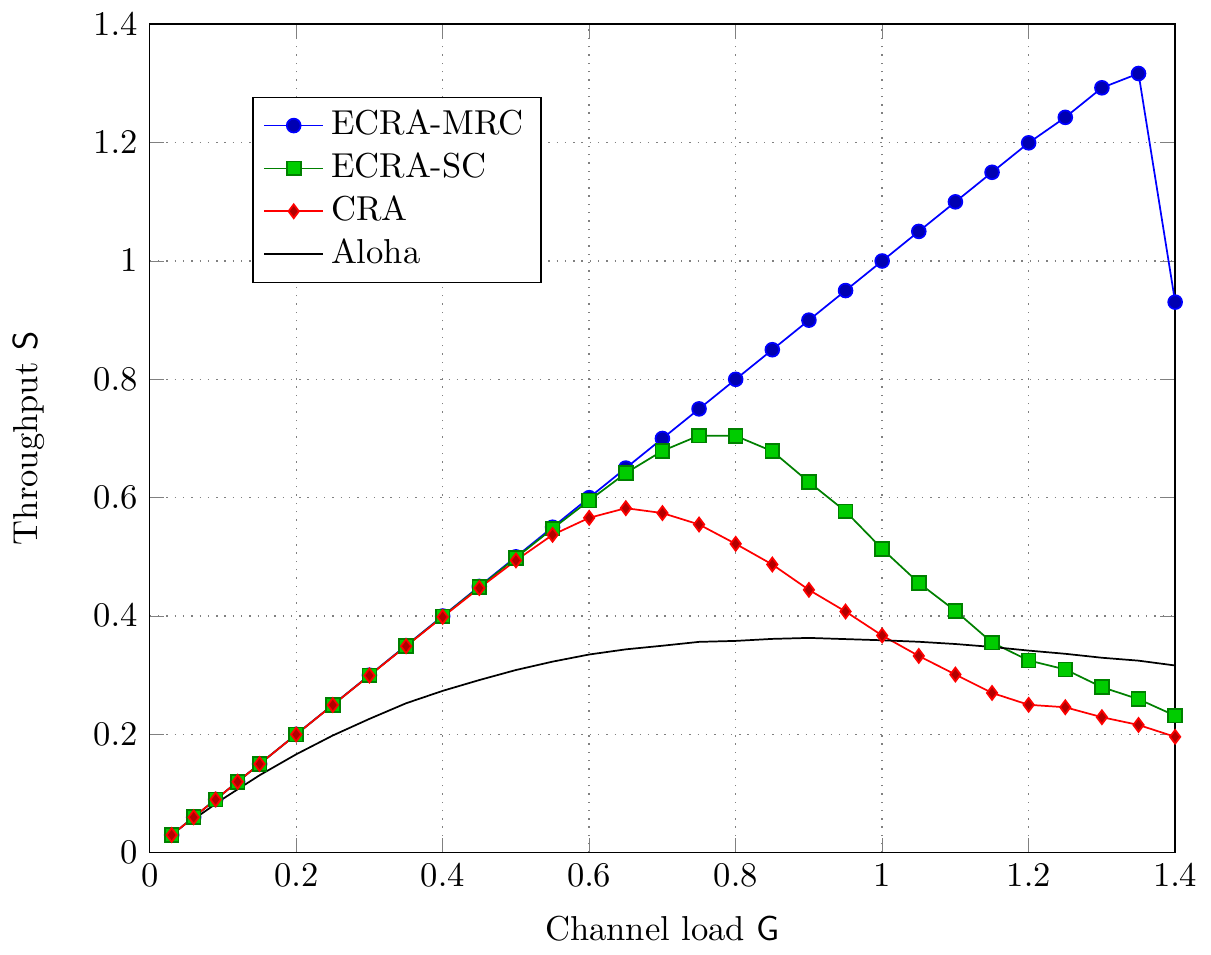}
\fi
\caption{Throughput $\tp$ vs. channel load $\load$ for ALOHA, \ac{CRA},
\ac{ECRA-SC} and \ac{ECRA-MRC}, $\usPw/\noisePw=6$ dB and $\rate=1.5$.}
\label{fig:T}
\end{figure}
In Fig. \ref{fig:T} the throughput $\tp$ vs. the channel load $\load$ is presented. \ac{ECRA-MRC} largely outperforms both \ac{ECRA-SC} and \ac{CRA}, reaching a maximum throughput of $\tp=1.32$ at $\load=1.35$, which is more than twice the one of \ac{CRA}, $\tp=0.58$ and $89$\% of
increase with respect to the one of \ac{ECRA-SC}, $\tp=0.70$. Furthermore, \ac{ECRA-MRC} throughput follows linearly the channel load up to $1.3$ packets per $\pkLen$, implying very small \ac{PLR}. In fact, looking at the \ac{PLR} performance in Fig. \ref{fig:PER},
\ac{ECRA-MRC} is able to maintain the \ac{PLR} below $10^{-3}$ for channel load below $1.2$ packets per $\pkLen$.
In other words for a target \ac{PLR} of $\plr=10^{-3}$, \ac{ECRA-MRC} can be operated up to $\load=1.2$, while both \ac{ECRA-SC} and \ac{CRA} only up to $\load\cong0.3$ and $\load\cong0.2$ respectively. The gain of \ac{ECRA-MRC} with respect to both \ac{ECRA-SC} and \ac{CRA} in terms of \ac{PLR} is of at least one order of magnitude, except in the very high channel load region, where it largely exceeds this value. It is also shown in the figure, that this protocol is the only one that can maintain $\plr\leq10^{-4}$ for channel load values up to $\load=0.6$. Very low \ac{PLR} are particularly appealing in specific scenarios as satellite applications or control channels where reliability can be as important as efficiency.

In Fig. \ref{fig:PER}, the approximation on the $\plr$ for both \ac{CRA} and \ac{ECRA-MRC}, derived in Section \ref{sec:Approx_PER}, is also shown. This approximation takes into account only the errors coming from \ac{UCP}s involving two users, and for very limited channel load values is very close to the simulated $\plr$. In particular, for the \ac{PLR} approximation of \ac{CRA} we used equation \eqref{eq:app_a_5} and equation \eqref{eq:plr_approx_final}, while for the \ac{PLR} approximation of \ac{ECRA-MRC} we used equations \eqref{eq:app_a_8} and \eqref{eq:plr_approx_final}, with the numerically evaluated average $\alpha$. For \ac{CRA}, when $\load\leq0.3$, the approximation approaches the $\plr$ simulated performance, while for increasing $\load$ the probability of having \ac{UCP}s involving more than two users starts to have an impact on $\plr$ and therefore the approximation starts to become loose. Although a similar behavior can be found for the approximation of \ac{ECRA-MRC}, interestingly the relative distance between the approximation and the simulations remains almost constant for a large range of channel load values.

In the second set of simulations, performance comparison of the slot synchronous scheme \ac{CRDSA} with the asynchronous schemes \ac{CRA}, \ac{ECRA-SC} and \ac{ECRA-MRC} is presented. The metrics used for the comparison are the spectral efficiency $\se$ and the packet loss rate $\plr$. We show numerical results for various rates. %We recall for the sake of completeness that \ac{CRDSA} has the same throughput $\tp$ performance for $\log_2 \left( 1 + \frac{\usPw}{\noisePw + \usPw}\right) < \rate \leq \log_2\left( 1 + \frac{\usPw}{\noisePw} \right)$, under the assumption of equal received power for all users and no multi-packet reception. Therefore, for \ac{CRDSA} in both simulations the rate $\rate^{\mathrm{s}}=\log_2\left( 1 + \frac{\usPw}{\noisePw} \right)$ is selected. For asynchronous schemes (\ac{CRA}, \ac{ECRA-SC} and \ac{ECRA-MRC}) instead, a rate $\rate^{\mathrm{a}}$ with $\rate^{\mathrm{a}}<\log_2\left( 1 + \frac{\usPw}{\noisePw} \right)$ is chosen.
We select rate $\rate=1.5$ for \ac{CRA} and \ac{ECRA-SC}, which is in line with the previous numerical results. Instead, for \ac{ECRA-MRC} we choose two different rate values, i.e. $\rate=1.5$ and $\rate=0.67$, where the latter is adopted to present the benefits of strengthening the error correction capabilities. Finally, for \ac{CRDSA}, we present results for $\rate=2.31$, which corresponds to the best choice when decoding in presence of one interferer is not possible, and for $\rate=0.67$, so to compare the performance at lower rate with \ac{ECRA-MRC}.
\begin{figure*}[t]
\begin{center}
\ifCLASSOPTIONtwocolumn
    \begin{subfigure}{0.85\columnwidth}
\else%
    \begin{subfigure}{0.5\columnwidth}
\fi
\centering\captionsetup{width=.95\textwidth}
\includegraphics[width=\columnwidth]{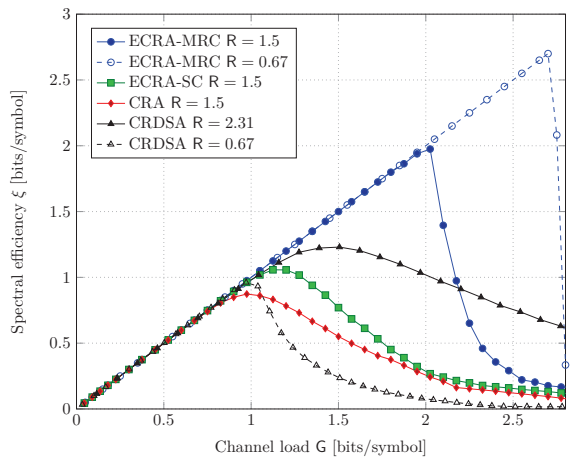}
   \caption{Spectral efficiency at $\usPw/\noisePw=6$ dB for various rate values.}
\end{subfigure}
\ifCLASSOPTIONtwocolumn
    \hspace{1.1cm}
\fi
\ifCLASSOPTIONtwocolumn
    \begin{subfigure}{0.85\columnwidth}
\else%
    \begin{subfigure}{0.5\columnwidth}
\fi
\centering\captionsetup{width=.95\textwidth}
\includegraphics[width=\columnwidth]{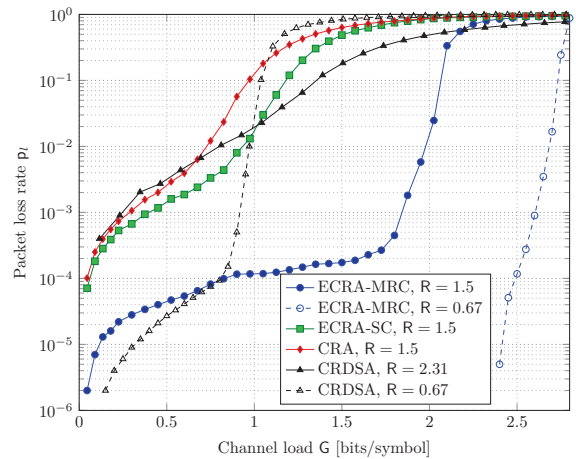}
   \caption{Packet loss rate at $\usPw/\noisePw=6$ dB for various rate values.}
   \label{fig:plr_sp_eff}
\end{subfigure}
\caption{\ $\se$ and \ac{PLR} vs. channel load $\load$ for \ac{ECRA-SC}, \ac{ECRA-MRC}, \ac{CRA} and \ac{CRDSA} for different rate values.}
\label{fig:sp_eff}
\end{center}
\end{figure*}
In Fig. \ref{fig:sp_eff}, the spectral efficiency results are presented. Considering \ac{CRDSA} with $\rate=2.31$ and \ac{CRA}, \ac{ECRA-SC} with $\rate=1.5$, we can observe that \ac{CRDSA} can outperform considerably both \ac{CRA} and \ac{ECRA-SC}. Specifically, it shows a peak throughput that is $16\%$ higher than \ac{ECRA-SC} and $40\%$ higher than \ac{CRA}. Nonetheless, comparing the \ac{PLR} results shown in Fig. \ref{fig:plr_sp_eff}, we observe a very similar performance for low channel load values, up to $\load=0.6$ $\mathrm{[bits/symbol]}$. On the other hand, when this channel load is exceeded, \ac{CRDSA} is able to gain on both \ac{CRA} and \ac{ECRA-SC}. The asynchronous \ac{ECRA-MRC} with $\rate=1.5$, instead, shows an outstanding gain of $60\%$ in the maximum spectral efficiency of \ac{ECRA-MRC} over \ac{CRDSA} with $\rate=2.31$, reaching a spectral efficiency close to $2$ $\mathrm{[bits/symbol]}$. Reducing the rate to $\rate=0.67$ leads to gains for both \ac{ECRA-MRC} and \ac{CRDSA}, looking at the \ac{PLR} performance. For the former, \ac{PLR} values below $10^{-6}$ are experienced for channel load up to $\load=2.4$ $\mathrm{[bits/symbol]}$, while for the latter it is achieved a \ac{PLR} below $10^{-4}$ up to $\load=0.8$ $\mathrm{[bits/symbol]}$ and below $10^{-3}$ up to $\load=0.9$ $\mathrm{[bits/symbol]}$. Comparing \ac{ECRA-MRC} for $\rate=1.5$ and for $\rate=0.67$ we observe a drastic improvement in both the \ac{PLR} performance as well as in the spectral efficiency when we select the lower rate. Although not presented in the figure due to space constraints, also for \ac{CRA} and \ac{ECRA-SC} a remarkable performance improvement can be shown for a rate of $\rate=0.67$ over $\rate=1.5$.

\begin{figure*}[t]
\begin{center}
\ifCLASSOPTIONtwocolumn
    \begin{subfigure}{0.85\columnwidth}
\else%
    \begin{subfigure}{0.5\columnwidth}
\fi
\centering\captionsetup{width=.9\textwidth}
\includegraphics[width=\columnwidth]{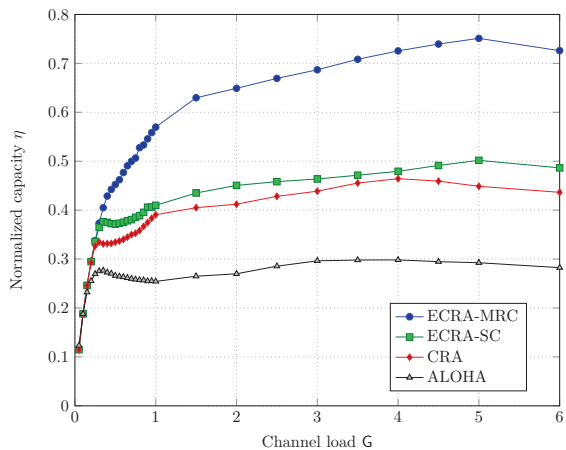}
\caption{Normalized capacity $\normCap$ for ALOHA, \ac{CRA}, \ac{ECRA-SC} and \ac{ECRA-MRC} with $\PwAg/\noisePw=6$ dB.}
   \label{fig:norm_cap}
\end{subfigure}
\ifCLASSOPTIONtwocolumn
    \hspace{1.1cm}
\fi
\ifCLASSOPTIONtwocolumn
    \begin{subfigure}{0.85\columnwidth}
\else%
    \begin{subfigure}{0.5\columnwidth}
\fi
\centering\captionsetup{width=.9\textwidth}
\includegraphics[width=\columnwidth]{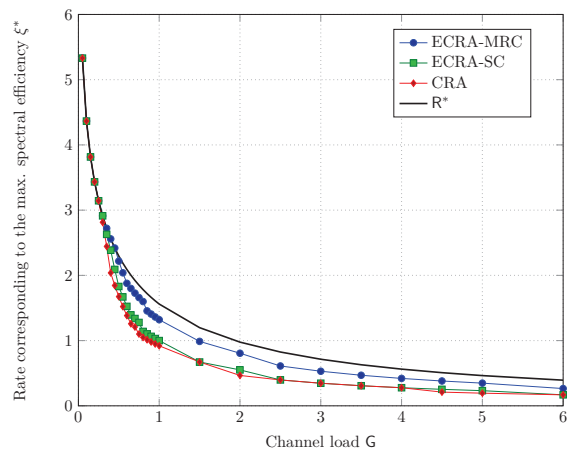}
\caption{Rate maximizing the spectral efficiency for \ac{CRA}, \ac{ECRA-SC} and \ac{ECRA-MRC} with $\PwAg/\noisePw=6$ dB.}
   \label{fig:rate_max}
\end{subfigure}
\caption{Normalized capacity $\normCap$ for ALOHA, \ac{CRA}, \ac{ECRA-SC} and \ac{ECRA-MRC} with $\PwAg/\noisePw=6$ dB and corresponding rate.}
\label{fig:norm_cap_tot}
\end{center}
\end{figure*}

The last set of simulations shows the comparison among ALOHA, \ac{CRA}, \ac{ECRA-SC} and \ac{ECRA-MRC}, in terms of the normalized capacity $\normCap$. $\PwAg/\noisePw=6$ dB is selected and the results are presented in Fig. \ref{fig:norm_cap_tot}.
%\begin{figure}
%\centering
%\subfigure[Normalized capacity $\normCap$ for ALOHA, \ac{CRA}, \ac{ECRA-SC} and \ac{ECRA-MRC} with $\PwAg/\noisePw=6$ dB.]{
%    \includegraphics[width=0.47\columnwidth]{ECRA_3_Norm_Eff/Norm_Eff}
%    \label{fig:norm_cap}
%}\hspace{0 mm}
%\subfigure[Rate maximizing the spectral efficiency for \ac{CRA}, \ac{ECRA-SC} and \ac{ECRA-MRC} with $\PwAg/\noisePw=6$ dB.]{
%    \includegraphics[width=0.47\columnwidth]{ECRA_3_Norm_Eff/R_max}
%    \label{fig:rate_max}
%    }
%\caption{Normalized capacity $\normCap$ for ALOHA, \ac{CRA}, \ac{ECRA-SC} and \ac{ECRA-MRC} with $\PwAg/\noisePw=6$ dB and corresponding rate.}
%\label{fig:norm_cap_tot}
%\end{figure}
The normalized capacity for \ac{ECRA-MRC} can reach up to 75\% of the \ac{MAC} channel capacity, for a channel load $\load=5$ with rate $\rate\cong0.35$; see Fig. \ref{fig:rate_max}. At this channel load, the gain is 50\% with respect to \ac{ECRA-SC} and 67\% with respect to \ac{CRA}. Interestingly, the normalized capacity for \ac{ECRA-MRC} as well as for both \ac{ECRA-SC} and \ac{CRA} is relatively constant for heavy channel load i.e., $\load>3$. In this way, the schemes appear to be robust against channel load fluctuations. On the other hand, the rate for which the maximum spectral efficiency $\maxSe$ (and so the normalized capacity) of the schemes is achieved lowers as the channel load increases; see Fig. \ref{fig:rate_max}. Therefore the system would be required to adapt the rate in order to reach the best performance in terms of normalized capacity. Nevertheless, the adaptation of the rate remains quite limited in this channel load region, ranging from a maximum of 0.53 at $\load=3$ to a minimum of 0.27 at $\load=6$ for \ac{ECRA-MRC}. For limited channel load, all the schemes performs very close, with ALOHA being slightly the best option. This is due to the low collision probability and the benefit of double transmit power of ALOHA compared to \ac{CRA} or \ac{ECRA} since no replicas are sent.

In Fig. \ref{fig:rate_max}, the rate corresponding to the maximum spectral efficiency for \ac{ECRA-MRC}, \ac{ECRA-SC} and \ac{CRA} is shown. The maximum possible rate under this scenario, is also depicted with a solid line in the figure. For limited channel load, the maximum spectral efficiency is achieved when using the maximum rate allowed, supporting the fact that collisions of received packets are seldom and the spectral efficiency can be maximized pushing the rate as much as it is allowed. On the other hand, as soon as the channel load exceeds $\load=0.3-0.4$, the maximum spectral efficiency is reached for rate values below the maximum one. In this way, the maximum spectral efficiency under moderate to high channel load conditions can be maximized taking a margin with respect to the maximum rate. This margin is helpful to counteract part of the collisions and at the same time does not reduces heavily the spectral efficiency.\footnote{Please note that the rate for ALOHA is not depicted in Fig. \ref{fig:rate_max} because it has a different degree $\dg$, and therefore the results are not comparable.} %As a final remark, for ALOHA, \ac{CRA} and \ac{ECRA-SC} when the rate maximizing the spectral efficiency moves away from the maximum rate, the maximum spectral efficiency experiences a local maximum, followed by a local minimum. For example, looking at the \ac{ECRA-SC} curves of the maximum normalized capacity, we can observe that a local maximum is found for $\load=0.35$, which corresponds to the last rate on the maximum allowed in Fig. \ref{fig:rate_max}, after this point the maximum spectral efficiency changes slope, and the rate drifts away from the maximum rate.

\section{Conclusions}
\label{sec:conclusions}
\acreset{UCP}

In this paper, a novel frame- and slot-asynchronous \ac{RA} decoding algorithm, named \ac{ECRA}, has been presented. Motivated by the presence of \acsp{UCP}, \ac{ECRA} tries to reduce their detrimental impact on the receiver's \ac{SIC} procedure, applying combing techniques. \ac{ECRA} exploits the presence of multiple instances of the same packet, in order to trigger the \ac{SIC} procedure. In addition, \ac{ECRA} tries to further reduce the interference, attempting to resolve partial collisions among packets, with the creation of a combined observation. The combined observation can be, either generated from the replicas sections with the lowest level of interference, resorting to \ac{SC}, or from the weighted combination of the replicas symbols of each user, resorting to \ac{MRC}. An analytical approximation of the \ac{PLR}, particularly tight for low to moderate channel load, is derived, considering only the \ac{UCP} involving two users. A comprehensive framework, with several metrics, is exploited for comparing both asynchronous and slot synchronous schemes, in the presence of channel coding. Finally, an investigation on the performance of \ac{ECRA} under average power constraint, is performed. Numerical simulations show that, \ac{ECRA} in both its variants, largely outperforms \ac{CRA}, for all the considered scenarios, in terms of both throughput and \ac{PLR}. Throughput exceeding $1$ packet per packet duration and \ac{PLR} below $10^{-4}$ for channel load up to $\load=0.6$ are achieved by \ac{ECRA-MRC}. Gains of up to $100$\% in the maximum throughput, w.r.t. \ac{CRA}, can be expected adopting \ac{ECRA-MRC} while \ac{ECRA-SC} has an improvement of 21\% w.r.t. \ac{CRA}. For a properly selected rate, \ac{ECRA-MRC} is also able to outperform \ac{CRDSA} with the same number of replicas. Finally, \ac{ECRA-MRC} shows remarkable performance gains in terms of normalized capacity w.r.t. the other asynchronous \ac{RA} schemes, reaching up to 75\% the \ac{MAC} channel capacity.

\section*{Acknowledgment}
The authors would like to thank Dr. Gianlugi Liva, German Aerospace Center (DLR), for the useful discussions.

\ifCLASSOPTIONcaptionsoff
  \newpage
\fi

\bibliographystyle{IEEEtran}
\bibliography{IEEEabrv,References}

\begin{biography}[{\includegraphics[width=1in,height=1.25in,clip,keepaspectratio] {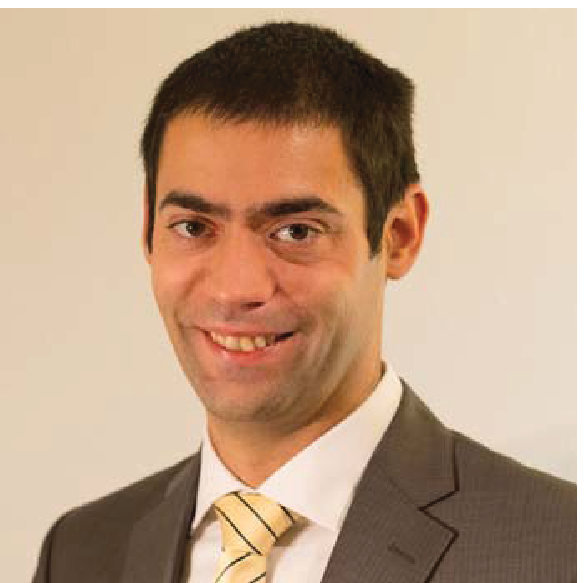}}]{Federico Clazzer} (S'11) was born in Genoa, Italy in 1987. He received the M.S. and the Ph.D. degrees in electrical engineering from the University of Genoa (Italy) in 2012 and 2017, respectively. His main research interests include satellite communication systems, random access techniques and signal processing algorithms. Since 2012 he is with the Institute of Communications and Navigation of the German Aerospace Center (DLR). During the past years he has been involved in several national and international projects on advanced medium access and random access techniques. In 2014, 2015 and 2016 he has been a frequent visitor at the Institute of Network Coding (INC), the Chinese University of Hong Kong. Aim of the collaboration has been the development of network coding techniques for the satellite communication scenario. Dr. Clazzer is IEEE member and he serves IEEE as reviewer for Transactions, Journals and Conferences as well as a technical program committee member. In 2017 he has been appointed as Exemplary Reviewer for IEEE Transactions on Communications.
\end{biography}

\begin{biography}[{\includegraphics[width=1in,height=1.25in,clip,keepaspectratio] {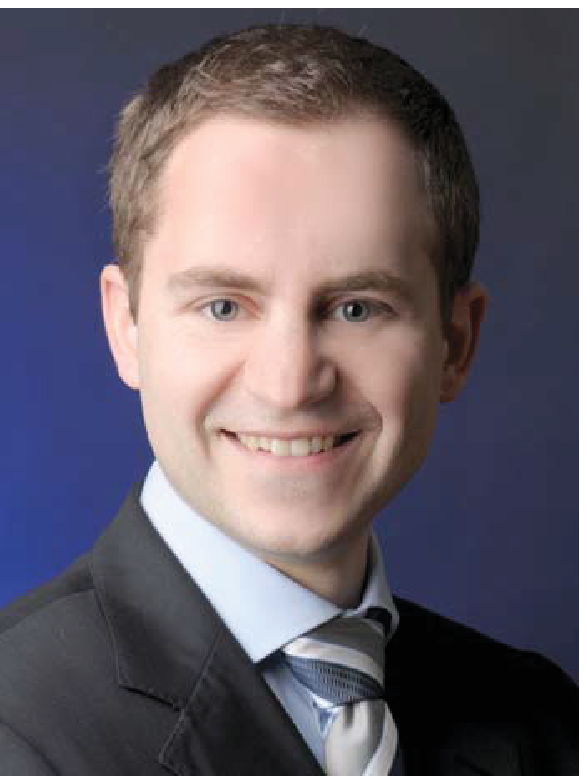}}]{Christian Kissling} Prof. Dr. Christian Kissling was born in Berchtesgaden, Germany in 1980. He received his Dipl.-Ing. degree in Electrical Engineering and Information Technology from the Technische Universit\"{a}t M\"{u}nchen (TUM) in 2005 and his Ph.D. degree with \emph{magna cum laude} in electrical engineering from the Technische Universit\"{a}t M\"{u}nchen (TUM) in 2014. In 2005 he joined the German Aerospace Center (DLR), Oberpfaffenhofen as a scientific researcher and has been working there at the institute of communication and navigation in the area of satellite and aeronautical communication.  In 2014 he joined the company Zodiac Aerospace / Zodiac Inflight Innovations and led a product development of aeronautical 3GPP server units as part of in-flight-entertainment. In 2016 he then became Professor at the University of Applied Sciences Munich in the area of electrical and computer engineering. His current research interests include wireless communication technologies as well as embedded microcontroller systems for smart homes and ambient assisted living.
\end{biography}

\begin{biography}[{\includegraphics[width=1in,height=1.25in,clip,keepaspectratio] {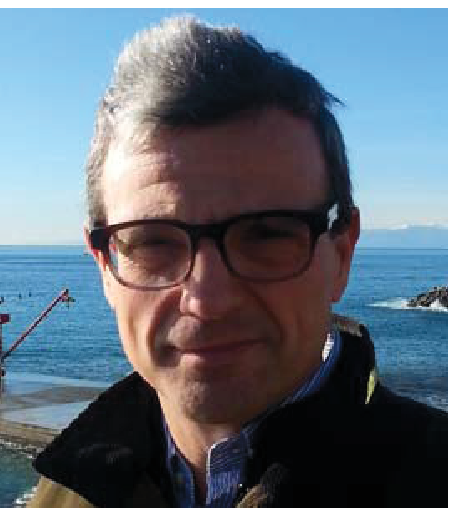}}]{Mario Marchese} (S'94-M'97-SM'04) was born in Genoa, Italy in 1967. He got his \emph{Laurea} degree cum laude at the University of Genoa, Italy in 1992, and his Ph.D. (Italian \emph{Dottorato di Ricerca}) degree in \emph{Telecommunications} at the University of Genoa in 1997. From 1999 to January 2005, he worked with the Italian Consortium of Telecommunications (CNIT), by the University of Genoa Research Unit, where he was Head of Research. From February 2005 to January 2016 he was Associate Professor at the University of Genoa. Since February 2016 he has been Full Professor at the University of Genoa. He was the Chair of the IEEE Satellite and Space Communications Technical Committee from 2006 to 2008. He is Winner of the IEEE ComSoc Award \emph{2008 Satellite Communications Distinguished Service Award} in \emph{recognition of significant professional standing and contributions in the field of satellite communications technology}. He is the author of the book \emph{Quality of Service over Heterogeneous Networks}, John Wiley \& Sons, Chichester, 2007, and author/co-author of more than 290 scientific works, including international magazines, international conferences and book chapters. His main research activity concerns: Networking, Quality of Service over Heterogeneous Networks, Software Defined Networking, Satellite DTN and Nanosatellite Networks, Networking security.
\end{biography}

\end{document}